\documentclass[aps,twocolumn,showpacs,groupedaddress]{revtex4}  
\usepackage{graphicx}  
\usepackage{epstopdf}
\usepackage{dcolumn}   
\usepackage{bm}        
\usepackage{amssymb}   
\usepackage{amsmath}
\usepackage{subfigure}
\usepackage{color}

\usepackage{float}

\begin{document}



\title{Nonequilibrium dynamics of a pure dry friction model subjected to coloured noise}
\author{Paul M. Geffert}
\author{Wolfram Just}

\affiliation{School of Mathematical Sciences, Queen Mary University of London, London E1 4NS, United Kingdom}                             

\begin{abstract}
We investigate the impact of noise on
a two-dimensional simple paradigmatic piecewise-smooth
dynamical system. For that purpose we consider the motion
of a particle subjected to dry friction and coloured noise. 
The finite correlation time of the noise provides an
additional dimension in phase space, causes a nontrivial probability
current, and establishes a proper nonequilibrium regime.
Furthermore, the setup allows for the study of 
stick-slip phenomena which show up as a singular component in the
stationary probability density. Analytic insight can be provided
by application of the unified coloured noise approximation, developed by 
Jung and H{\"a}nggi \cite{JUN87}. The analysis of probability currents
and of power spectral densities underpins the
observed stick-slip transition which is related with a critical
value of the noise correlation time. 
\end{abstract}

\pacs{}
\maketitle

\section{\label{sec:1}Introduction}
Piecewise-smooth dynamical systems have attracted a lot of interest in the last decade. They are widely used to model switching or impact behaviour in many different areas of science i.e. as biology, engineering, physics or mathematics \cite{DIB08,FIL88,MAK12,COO12,JEF11}. 
Systems with dry (or Coulomb) friction are prominent examples in the context of piecewise-smooth models \cite{BIE12}. The main feature of this type of friction is that an applied force has to overcome a certain threshold to move an object (sliding), otherwise the object rests (sticking) \cite{BOW50}. This behaviour is usually modelled by a sign-function and allows a simple macroscopic description for systems where solid-solid interactions are important, e.g. as for stick-slip dynamics \cite{ELM97,WEN06}. 
Adding noise to the dynamical equations of a piecewise-smooth system opens a whole new area of research, which is still in its infancy. The interplay of dry friction and random forces has been reported in \cite{KAW04,DEG05,HAY05}. Exact solutions are known for a few piecewise-smooth stochastic models, where e.g. the propagator can be obtained for the case of pure dry friction \cite{CAU61,TOU10} or in connection with Laplace transforms \cite{TOU12}. Other analytical results are available in the framework of path integrals and weak noise approximations \cite{BAU10, BAU11, CHE13} or first passage time problems \cite{CHE14}.
Whereas the aforementioned studies are dealing with Gaussian white noise, models with
Non-Gaussian noise and dry friction have been investigated as well \cite{BAU12,BAU13, KAN15}.
The features of systems with dry friction subjected to random forces have also been observed in experimental setups \cite{CHA08,GOO09, GOO10,GNO13, GNO13a}. 
From a more rigorous mathematical point of view, the impact of a stochastic perturbation on a piecewise-smooth dynamical system has been considered in \cite{SIM14,SIM15}.

A profound understanding of the impact of noise on piecewise-smooth dynamical systems is desirable from an intrinsic
theoretical perspective, and will contribute as well to relevant experimental issues. For instance, nonequilibrium properties of granular
media is a topical subject, see e.g. \cite{ESH10} for recent experimental results. The corresponding theoretical modelling uses granular material as a nonequilibrium heat bath, and studies the impact on devices subjected to dry friction.
Localisation phenomena of the velocity and intermittent dynamics are consequences of the underlying stick-slip dynamics \cite{TAL11,TAL12,SAN16}.
Realistic theoretical models are fairly complicated so that only very 
limited analytic insight can be obtained.

The inclusion of a finite correlation time of the noise is
a simple way to emulate nonequilibrium properties of a heat bath. If the correlation time of the noise is 
of the same order as the characteristic time scale of the system a
correlated noise (or coloured noise) is required 
\cite{HAE95}. Analytical treatments of coloured noise
are hampered by the lack of detailed balance. 
In this context the so-called unified coloured noise approximation (UCNA) has been developed by Jung and H{\"a}nggi to obtain analytic expressions for the stationary probability density \cite{JUN87}. Coloured noise has been studied in many different contexts, e.g.  magnetic resonance systems \cite{KUB62} or neurodynamics \cite{BRA10}.

The purpose of our contribution is twofold. We want to investigate the impact of noise on piecewise-smooth dynamical systems in a simple setup which allows for a partial analytic treatment. Furthermore, the nonequilibrium aspects, the occurrence of stationary probability currents, and transition phenomena will be a crucial part of our investigations. This paper is organized as follows:
Section \ref{sec:2} introduces the pure dry friction model subjected to 
coloured noise. The stationary behaviour of the model is investigated
in Section \ref{sec:3}. Analytic expressions for the velocity
distribution will be derived together with an asymptotic expression
for the two-dimensional stationary distribution. Analytic results
are supported by numerical simulations for the density and for 
the stationary probability current. Dynamical properties such as 
the power spectral density and the distributions of sliding and sticking 
events are elaborated on in Section \ref{sec:4}. 
We conclude our studies in Section \ref{sec:5}.
\section{\label{sec:2}The model}

\begin{figure}[h]
\centerline{\includegraphics[width=0.43 \textwidth]{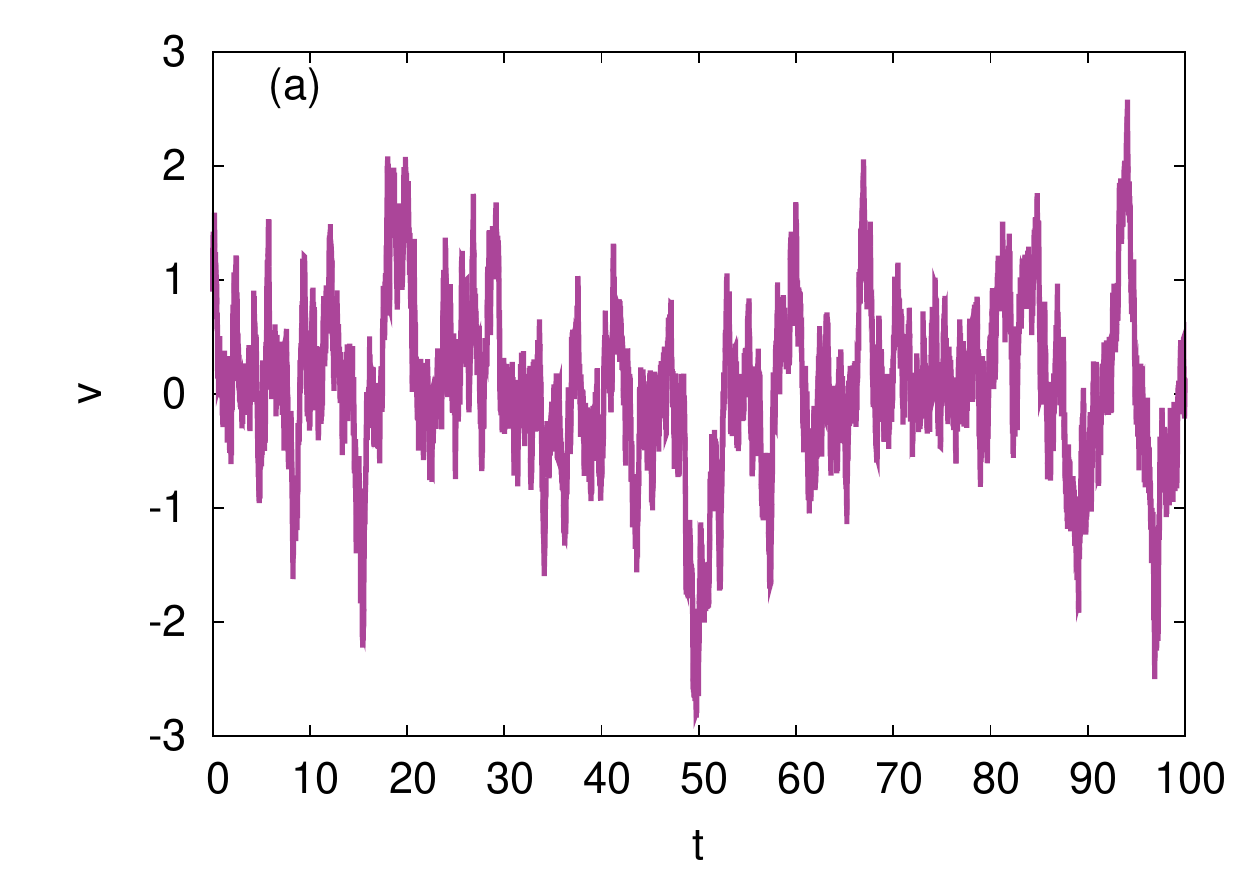}}
\centerline{\includegraphics[width=0.43 \textwidth]{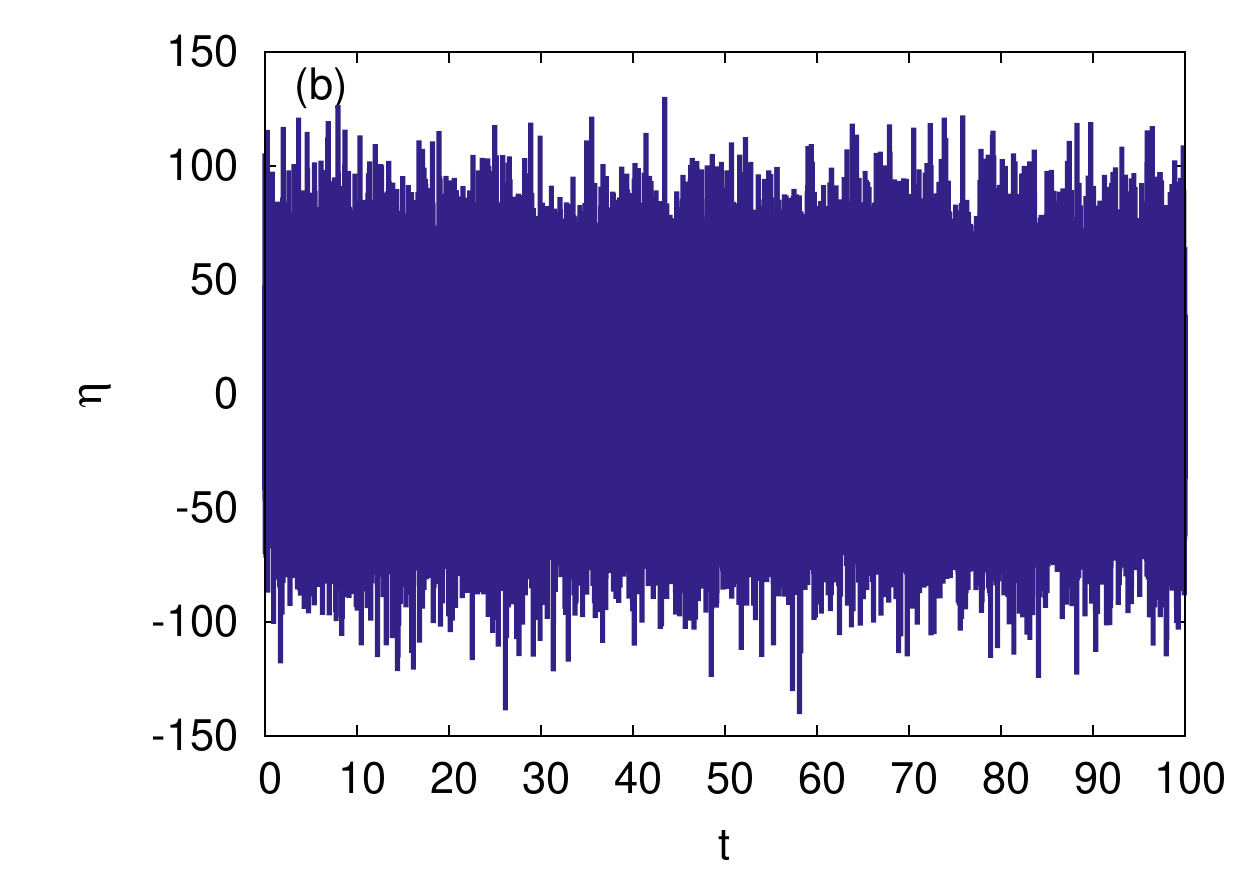}}
\caption{Time traces of the velocity $v(t)$ (a) and the Ornstein-Uhlenbeck noise $\eta(t)$ (b) of the dry friction model (eqs.(\ref{eq:2b}) and (\ref{eq:2c})) for $\tau=0.001$.}
\label{fig:2a}
\end{figure}
\begin{figure}[h]
\centerline{\includegraphics[width=0.43 \textwidth]{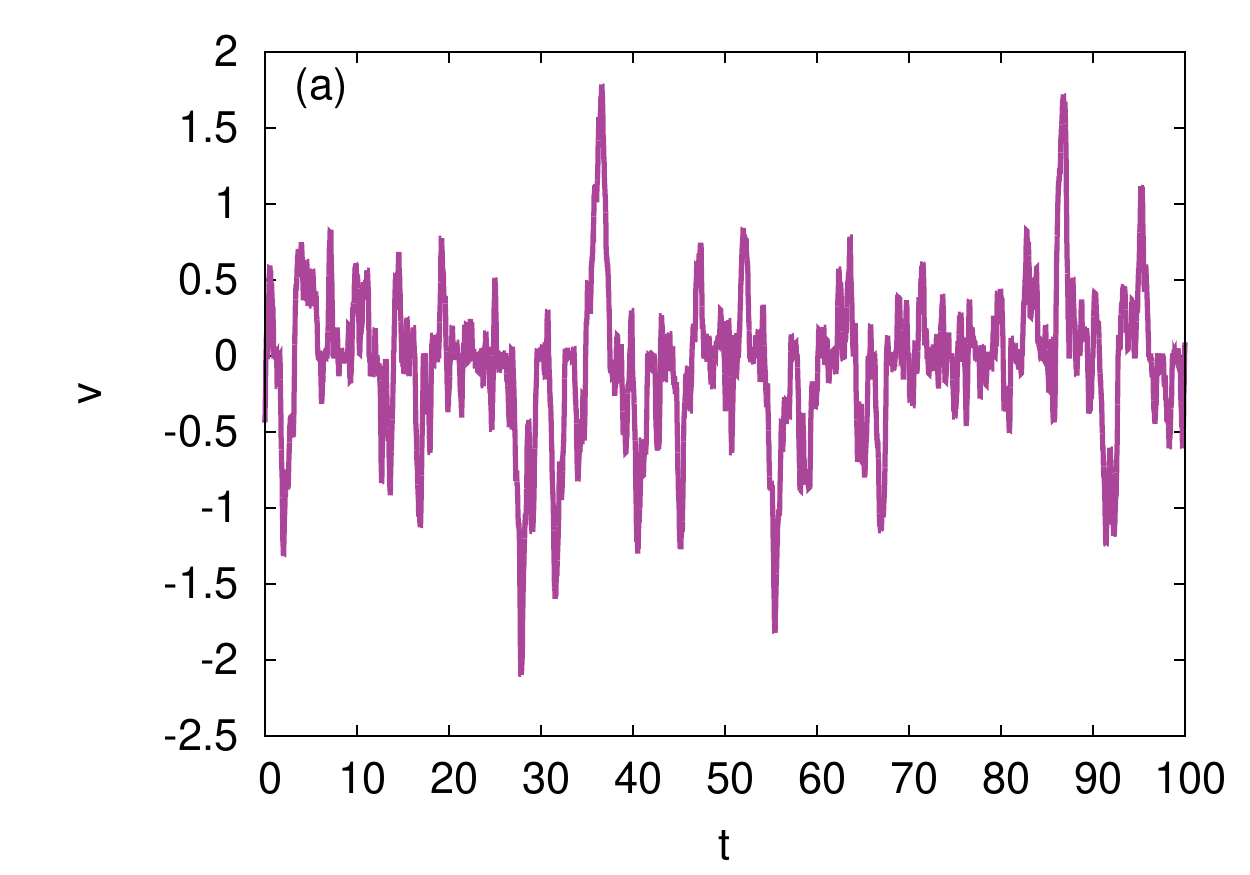}}
\centerline{\includegraphics[width=0.43 \textwidth]{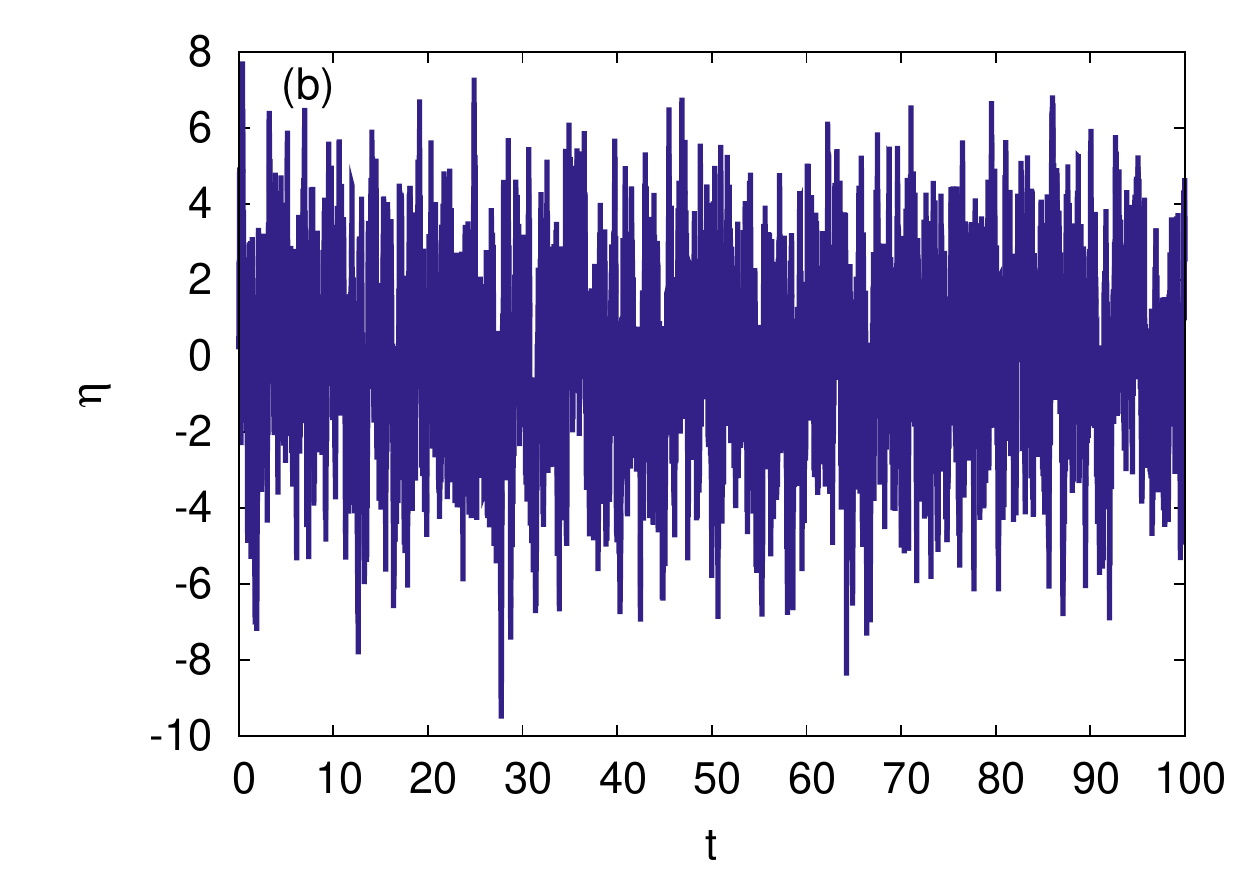}}
\caption{Time traces of the velocity $v(t)$ (a) and the Ornstein-Uhlenbeck noise $\eta(t)$ (b) of the dry friction model (eqs.(\ref{eq:2b}) and (\ref{eq:2c})) for $\tau=0.1$.}
\label{fig:2b}
\end{figure}
\begin{figure}[h]
\centerline{\includegraphics[width=0.43 \textwidth]{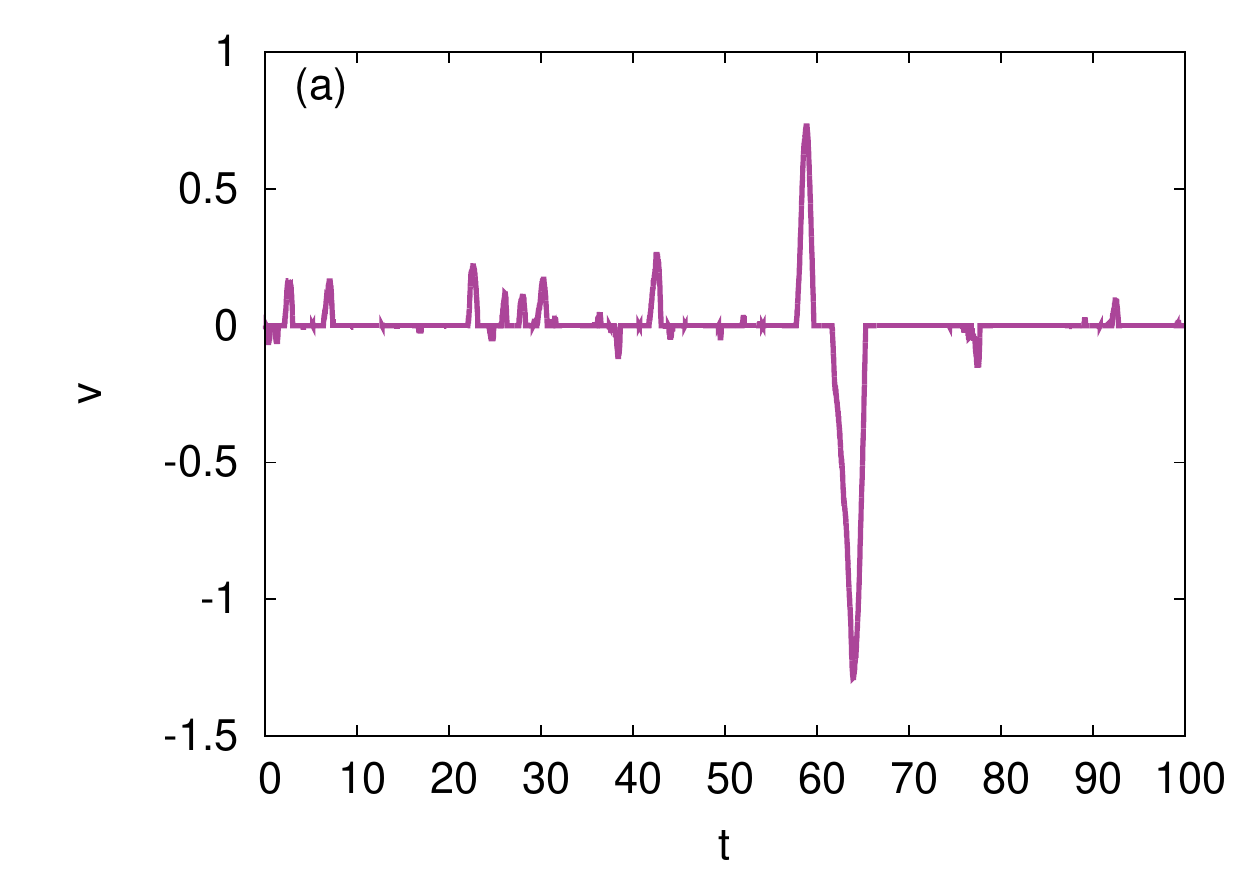}}
\centerline{\includegraphics[width=0.43 \textwidth]{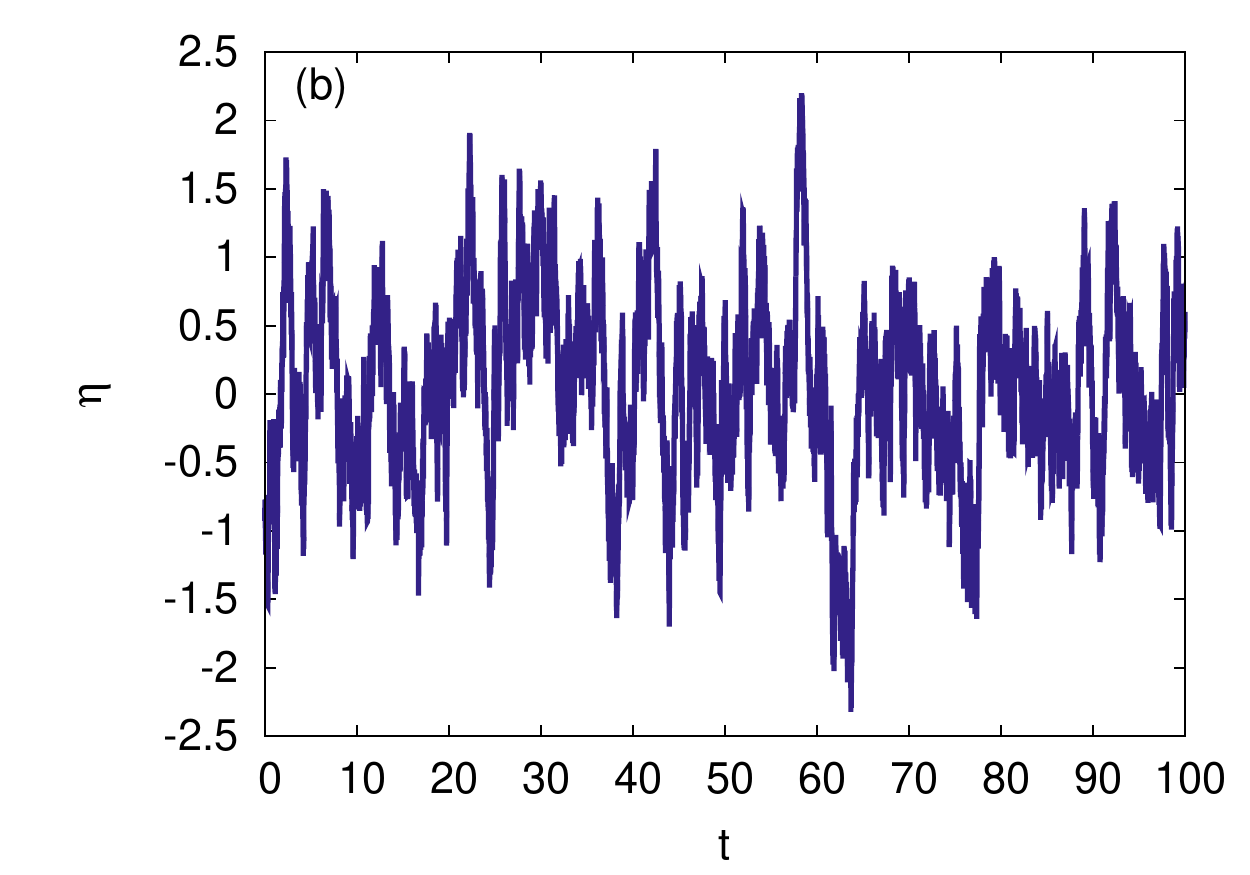}}
\caption{Time traces of the velocity $v(t)$ (a) and the Ornstein-Uhlenbeck noise $\eta(t)$ (b) of the dry friction model (eqs.(\ref{eq:2b}) and (\ref{eq:2c})) for $\tau=1.0$.}
\label{fig:2b1}
\end{figure}

We consider the simplest two-dimensional case of a 
piecewise-smooth stochastic system, which does not obey
detailed balance. To motivate our considerations let us
recall the one-dimensional motion of a particle subjected
to white noise. With a slight abuse of notation the corresponding
Langevin equation governing the velocity reads
\begin{equation}\label{eq:2a}
\dot{v}(t)=- \sigma_0(v(t)) + \xi(t), 
\end{equation}
where $\xi(t)$ denotes a white Gaussian noise with correlation
function $\langle \xi (t) \xi (s) \rangle = \delta (t-s)$ 
and
$\sigma_0(v)=\mbox{sign}(v)$ contains the deterministic part caused by
Coulomb friction. We have adopted units such that the noise intensity 
and the dry friction coefficient have been normalised to one. Eq.(\ref{eq:2a}) is not well defined at $v=0$. One could
cure such an inconsistency by considering the Coulomb friction as the
limiting case of the regular drift $\sigma_\varepsilon(v)=\tanh(v/\varepsilon)$
for $\varepsilon \rightarrow 0$. Such niceties are not relevant for
eq.(\ref{eq:2a}) as the white noise is not a function 
with well defined finite values and the stochastic model in a strict sense 
is not pointwise defined. The formally written down 
Fokker-Planck equation with suitable matching conditions ensuring continuity
of the density and continuity of the probability current captures
all aspects of the dynamics and has been studied intensely
in the literature, see, e.g., \cite{DEG05}. The deterministic part 
without noise requires a more careful approach in terms of piecewise-smooth 
dynamical systems \cite{DIB08}, in particular, in the presence 
of a finite amplitude driving force where stick-slip transitions 
occur (cf. eq.(\ref{eq:2c})).

An obvious extension of the model described above, leading towards a
two-dimensional stochastic nonequilibrium system, consists in studying
the effect of coloured noise. To be precise we intend to replace the 
Gaussian white noise by an exponentially
correlated Ornstein-Uhlenbeck process $\eta(t)$,
which is governed by the stochastic differential equation
\begin{equation}\label{eq:2b}
\dot{\eta}(t)=-\frac{\eta(t)}{\tau} + \frac{\xi(t)}{\tau} \, .
\end{equation}
The noise correlation time $\tau$ will be the only effective parameter in our
model. Since the process $\eta(t)$ can be viewed as a continuous
function some care is needed when introducing the dynamics of the
particle. For forces smaller than the dry friction coefficient,
$|\eta|<1$, and $v=0$ the particle will stick while otherwise the sliding 
dynamics is still described by the aforementioned equation of motion.\\ 
Thus we end up with
\begin{equation}\label{eq:2c}
\dot{v}(t) = \left\{
\begin{array}{ll} 0 & \mbox{ if } v=0 \mbox{ and } |\eta(t)|<1\\
-\sigma_0(v(t)) + \eta(t) &  \mbox{ otherwise} \end{array} \right. \, .
\end{equation}
Alternatively we could use the regularized drift
\begin{equation}\label{eq:2d}
\dot{v}(t) = -\sigma_\varepsilon(v(t)) + \eta(t)
\end{equation}
and consider results finally in the limit $\varepsilon\rightarrow 0$. We will adopt both views throughout our exposition while we will,
at the same time, avoid the considerable technical difficulties that 
would be related  with a rigorous approach. Since we consider a noise 
with finite correlation time but a damping which does not involve a
memory kernel, the system violates 
detailed balance and describes a nonequilibrium process \cite{KUB91}.
On the contrary, the model in the white noise limit, 
eq.(\ref{eq:2a}),
has a vanishing stationary probability current and describes a
process in equilibrium.

For the model defined by eq.(\ref{eq:2c}) the static friction
equals the kinetic friction. In real world the former exceeds the latter, and our assumption has to be considered as an idealisation from
an experimental perspective. 
Our particular choice does not include any hysteresis, but it has the advantage that
the piecewise-smooth dynamical system can be captured as a singular limit
of the smooth dynamics eq.(\ref{eq:2d}). Eq.(\ref{eq:2c}) provides the
simplest consistent version of a piecewise-smooth dynamical system
with a sliding region and an invisible tangent point \cite{DIB08}. 
Above all the model captures stick-slip phenomena which will be a key ingredient of our analysis.

Before we enter a more detailed discussion let us just illustrate the 
main phenomenon by time traces obtained from numerical simulations.
Throughout all our numerical investigations
we apply an Euler-Maruyama scheme with step size $h=10^{-3}$ for different 
values  of $\tau $. To take the
discontinuity caused by dry friction into account 
(see eq.(\ref{eq:2c})) we set $v = 0$ for $|v| < 10^{-3}$ and $|\eta| < 1$, 
as the particle sticks in this case at the origin. Time traces 
from the simulations are shown in figures \ref{fig:2a} - \ref{fig:2b1}. 
At a scale of order one the effect of dry friction becomes visible for 
correlation times larger than $\tau = 0.1$.  The particle sticks 
for considerable amounts of time at $v=0$, as the stochastic force 
$\eta(t)$ is not large enough to move the particle. It is this stick-slip phenomenon and the related intermittent motion which will be 
at the centre of our studies, being the key signature of our piecewise-smooth 
stochastic model.

The observed dynamics from the simulations seems to be a key feature of dry
friction subjected to noise and of general piecewise-smooth stochastic dynamics. Signatures of such intermittent dynamics have been found in the framework of the Boltzmann-Lorentz equation by investigating the so-called independent kick model \cite{TAL11,TAL12}, in studies of dry friction subjected to Non-Gaussian noise in the high friction limit \cite{KAN15}, and in an experiment of a rotating probe subjected to a granular gas \cite{GNO13a}. Intermittent dynamics and a related splitting of the velocity distribution in a regular and a singular
part can be clearly seen in numerical studies of the underlying
transport equations \cite{SAN16}. Despite the importance of dry friction in engineering, only few explicit results on its interplay with noisy nonequilibrium environments
are available in the literature. We think, that
justifies a case study like eq.(\ref{eq:2c})
to uncover potentially general features caused by
discontinuities of the flow and noise with finite amplitude.
Furthermore, the noise correlation time $\tau$ is used as a continuous control parameter in the analysis of our model. Such a parameter has not been available in the studies using the Boltzmann equation, where only limiting cases of frequent and rare collisions were investigated \cite{GNO13,GNO13a,TAL11,TAL12}.

\section{\label{sec:3}Stationary density}

Given the previous reasoning and the numerical findings we expect the stationary 
density to exhibit a singular component caused by particles sticking at
$v=0$. The corresponding stationary distribution is expected to consist 
of a Dirac $\delta$ component at vanishing velocities and $|\eta|<1$,
and a regular part describing moving particles with finite velocities. 
The analysis will be further hampered by the lack of detailed balance so 
that closed form analytic expressions are unlikely to be available.

\subsection{\label{sec:3a}Marginal distribution and 
unified coloured noise approximation}

To make some analytical headway let us first have a
look at the marginal velocity distribution $P_v(v)=\int_{-\infty}^{\infty}
P(v,\eta) d \eta$ for which perturbative treatments in terms of the
correlation time are available. We are interested in possible changes
compared to the white noise case $\tau \neq 0$
(see e.g. \cite{TOU10}). We apply the unified coloured noise 
approximation (UCNA), developed by Jung and H{\"a}nggi \cite{JUN87}, 
to our regularized system eqs.(\ref{eq:2b}) and (\ref{eq:2d}). This method can be seen 
as a kind of interpolation scheme for systems with coloured noise, as this 
method shows, under certain conditions, exact results in the limit of 
vanishing correlation 
$\tau \rightarrow 0$ and high correlation $\tau \rightarrow \infty$.

For the convenience of the reader, we recall the 
main steps of the derivation of the stationary probability density $P_v(v)$. 
If we eliminate the variable $\eta$ from eqs.(\ref{eq:2b}) and (\ref{eq:2d})
we obtain the second order equation
\begin{equation}\label{eq:2f}
\ddot{v}(t) + \dot{v}(t) \left(\sigma'_\varepsilon(v(t)) + 
\frac{1}{\tau}\right)=  -\frac{\sigma_\varepsilon(v(t))}{\tau} + 
\frac{1}{\tau}\xi(t) \, .
\end{equation}
We introduce a new time scale $\hat{t} = \tau ^{-1/2}t$,
\begin{equation}\label{eq:2g}
\ddot{v}(\hat{t}) + \dot{v}(\hat{t})
\gamma (v(\hat{t}), \tau) = -\sigma_\varepsilon (v(\hat{t})) + 
\tau ^{-1/4}\xi (\hat{t}),
\end{equation}
where we have the damping factor
\begin{equation}\label{eq:2h}
\gamma (v, \tau) = \tau ^{-1/2}+\tau ^{1/2}\sigma'_\varepsilon(v).
\end{equation} 
This factor approaches infinity for both limits $\tau \rightarrow 0$ 
and $\tau \rightarrow \infty$. Hence, the setup is suitable for
an adiabatic elimination scheme in the limit of small correlation times.
If we neglect the second order derivative
we obtain a simpler multiplicative stochastic process
\begin{equation}\label{eq:2i}
\dot{v}(\hat{t}) = 
-\frac{\sigma_\varepsilon(v(\hat{t}))}{\gamma (v(\hat{t}), \tau)} 
+ \frac{1}{\tau ^{1/4}\gamma (v(\hat{t}), \tau)}\xi (\hat{t})
\end{equation}
with a corresponding Fokker-Planck equation in the Stratonovich sense
\begin{eqnarray}\label{eq:2j}
\partial_{\hat{t}} P_v &=& \partial_v \left(\frac{\sigma_\varepsilon(v)}{
\gamma (v, \tau)} + \frac{1}{2 \tau ^{1/2}}
\frac{\gamma '(v, \tau)}{\gamma ^3(v, \tau)} \right)P_v \nonumber \\
& &
+ \frac{1}{2\tau ^{1/2}}
\partial _v^2\left(\frac{P_v}{\gamma ^{2}(v,\tau)}\right) \, .
\end{eqnarray}
Since the adiabatic approximation has reduced the problem to a one-dimensional
Fokker Planck equation the stationary distribution can be computed by
straightforward integration
\begin{equation}\label{eq:2k}
P_v(v) = 
\exp \left(-2 \int \sigma_\varepsilon(v)dv - \tau \sigma_\varepsilon^2(v)  
+ \ln \left(|1+\tau \sigma'_\varepsilon(v) |\right) \right).
\end{equation}
In the dry friction limit
$\varepsilon\rightarrow 0$
the normalised stationary probability density reads
\begin{equation}\label{eq:2m}
P_v(v) = \frac{\exp \left(-2|v| - \tau \sigma_0^2(v) \right)\left(1+\tau \delta (v) \right)}{\exp \left(-\tau\right)+\sqrt{\pi \tau} \text{erf} (\sqrt{\tau})} .
\end{equation}

Eq.(\ref{eq:2m}) shows that the stationary probability density consists 
of two  parts, a regular contribution for $v \neq 0$ and a singular part 
for $v = 0$. The delta contribution in the density reflects the fact that 
the particle sticks at $v=0$ when the stochastic force is not large enough 
to move the particle. The regular part of the density 
describes the sliding motion of the particle for $v \neq 0$. 
By taking the white noise limit $\tau \rightarrow 0$ we arrive at the 
exact stationary probability density for dry friction with white noise 
(i.e. \cite{TOU10}). 
For high correlation times $\tau$ the sliding contribution decreases 
and the density is mainly determined by the delta peak. Thus, by increasing 
$\tau$ we can observe a gradual transition from sliding 
to sticking motion.
The appearance of a delta peak in the expression for the stationary probability density has also been found in various theoretical studies \cite{TAL11,TAL12,BAU12,BAU13,KAN15,SAN16} and in experiments \cite{GNO13a}.

\begin{figure}[h]
\centerline{\includegraphics[width=0.43 \textwidth]{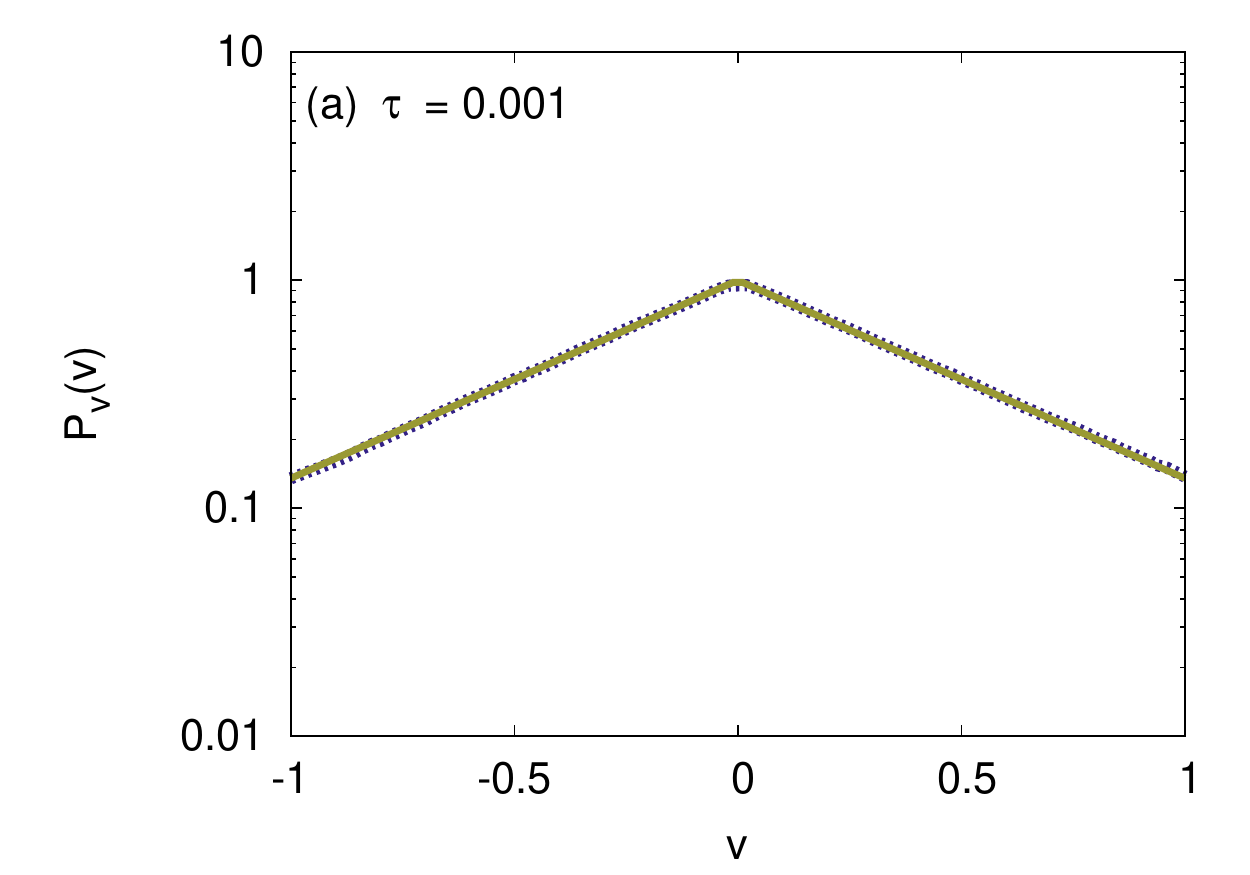}}
\centerline{\includegraphics[width=0.43 \textwidth]{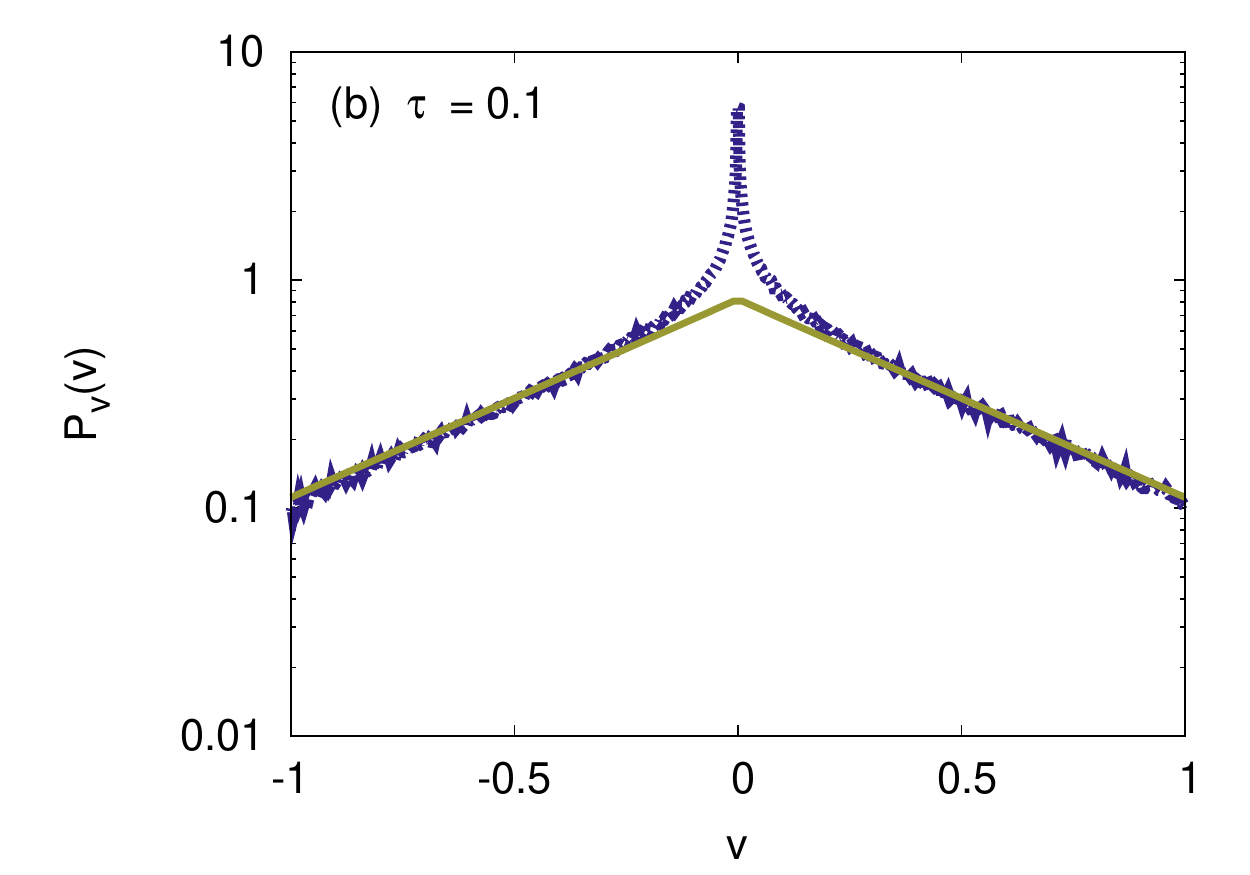}}
\centerline{\includegraphics[width=0.43 \textwidth]{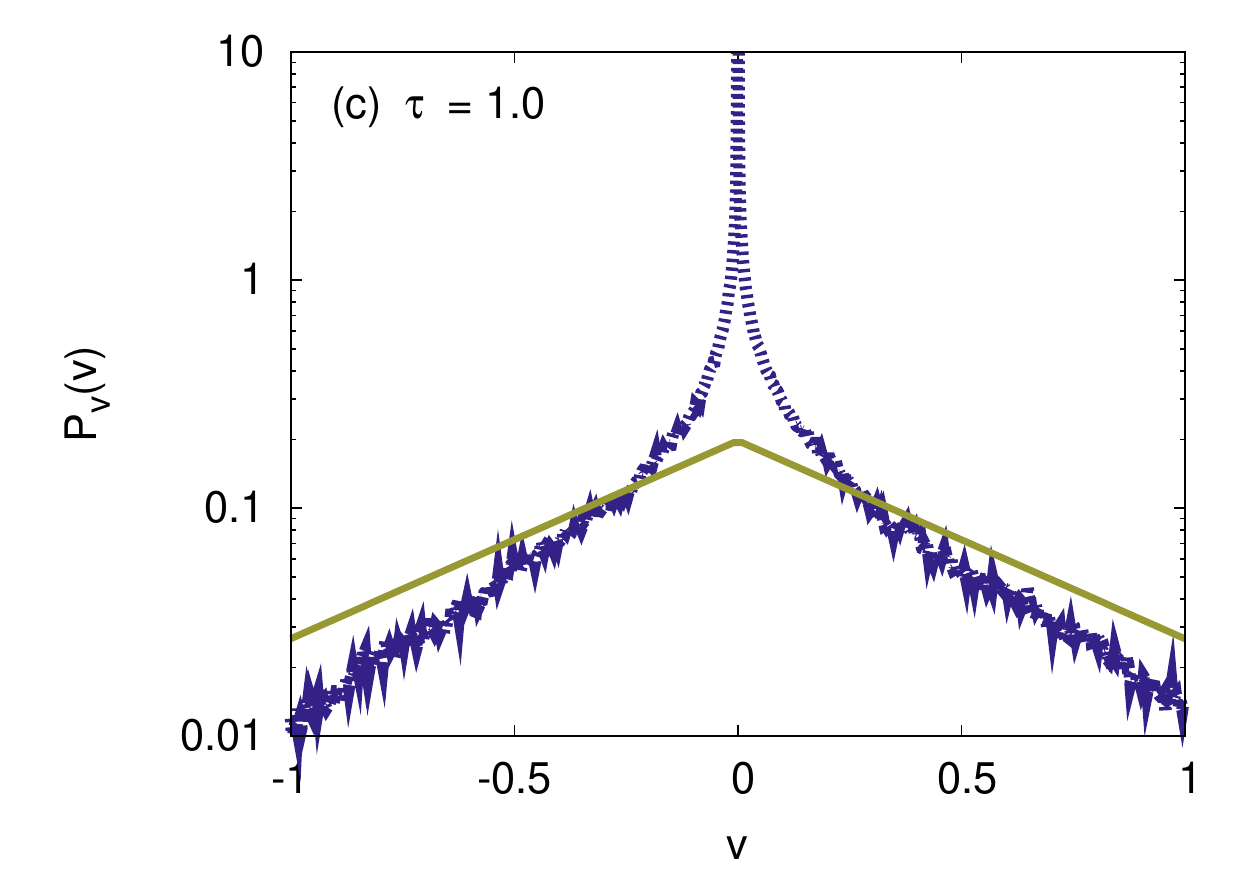}}
\caption{Regular part of the stationary density, i.e. distribution of the
sliding events, obtained from numerical simulations (dashed lines) sampled as a histogram
with resolution $\Delta v = 0.002$ and the analytical approximation, 
eq.(\ref{eq:2m}) (solid lines). Data have been displayed for different values
of the correlation time $\tau=0.001$ (a), $\tau=0.1$ (b), $\tau=1.0$ (c), cf. figures \ref{fig:2a} - \ref{fig:2b1}.}
\label{fig:2c}
\end{figure}

The accuracy of the perturbative approach can be confirmed by 
direct numerical simulations, see figure \ref{fig:2c} for the
comparison of the UCNA with direct numerical simulations.
By taking at about 100 realisations of 
time traces of length $T = 10^4$, we observe good agreement
for small correlation times. 
However, for values $\tau > 0.1$, deviations between numerics and analytics 
become visible.

\begin{figure}[h]
\includegraphics[width=0.43\textwidth]{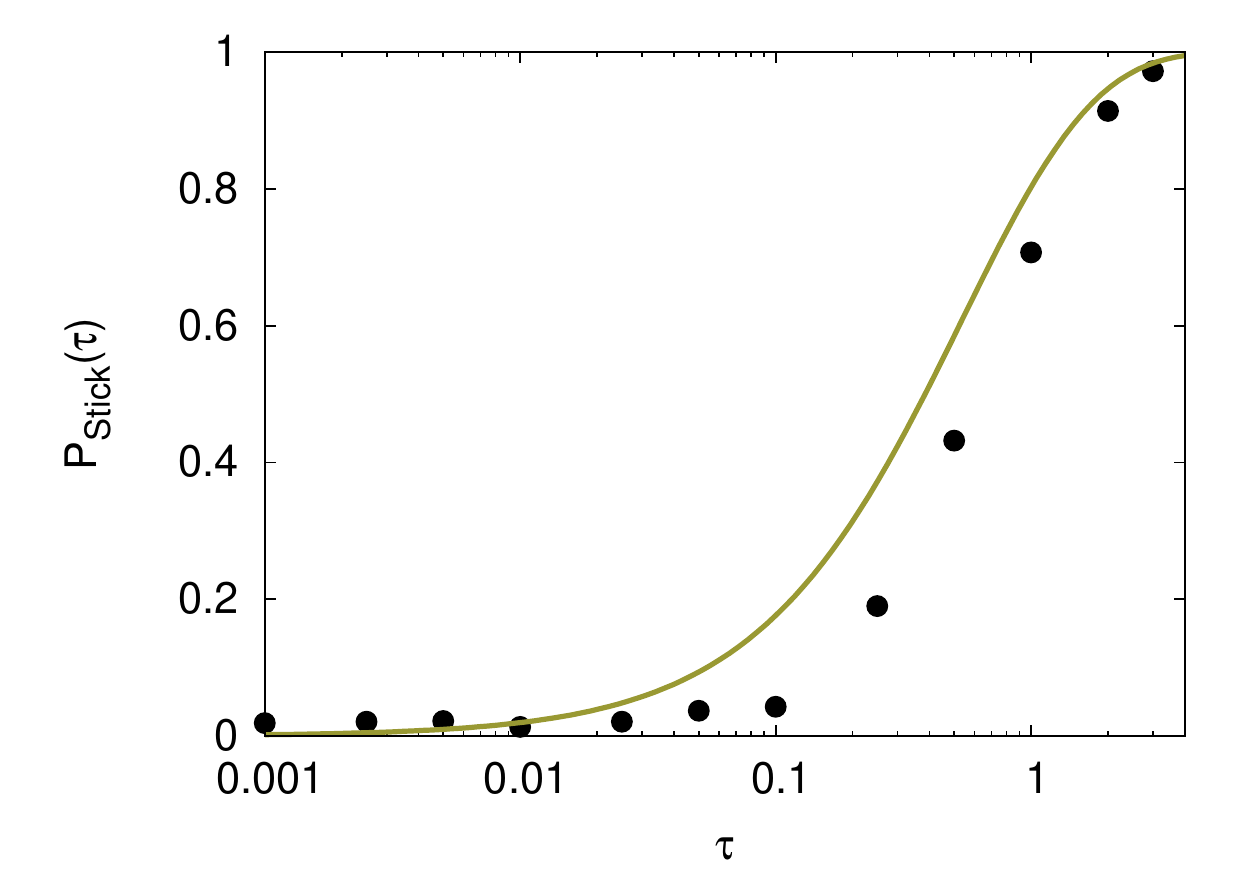}
\caption{Probability of the sticking events as a function of the noise correlation time $\tau$. The dots correspond to numerical simulations of eqs.(\ref{eq:2b}) and (\ref{eq:2c}), the solid line corresponds to eq.(\ref{eq:2m1}).}
\label{fig:2d}
\end{figure}

In addition to the analysis of sliding events, eqs.(\ref{eq:2k}) and (\ref{eq:2m}) 
give as well an estimate for the singular part, in particular for the probability of sticking as a function of the noise correlation
\begin{equation}\label{eq:2m1}
P_{Stick}(\tau) = \frac{\sqrt{\pi \tau} \text{erf} (\sqrt{\tau})}{\exp \left(-\tau\right)+\sqrt{\pi \tau} \text{erf} (\sqrt{\tau})}.
\end{equation}
To obtain this result one needs to integrate the regularized version,
eq.(\ref{eq:2k}), over a small interval containing $v=0$ and then taking
the limit $\varepsilon\rightarrow 0$.
Figure \ref{fig:2d} shows the comparison of the analytical approximation with the simulations and we observe a quite good agreement as the probability of sticking increases with increasing $\tau$ and approaches the value $1$ in the limit of high correlation times.

Overall, the analytic approximation seems to work rather well, especially for small $\tau$. Deviations become visible when the noise correlation time increases (see figure \ref{fig:2c} for the case $\tau = 1.0$).
To explain the deviations between the analytical approximation and the
direct numerical simulations for the regular part/sliding events (figure \ref{fig:2c}), we need to take a look at the conditions of 
validity of the UCNA; this approximation gives proper results for the 
case $\gamma (v,\tau) \gg 1$. But for higher values of $\tau$, this 
approximation fails as in our case the contribution caused by the dry friction
vanishes in the limit $\varepsilon \rightarrow 0$, i.e., when considering the
piecewise linear case. Nevertheless, this analytic approximation scheme provides very useful information of the underlying nonequilibrium dynamics of our model.

\subsection{\label{sec:3b}Joint distribution and probability current}

To get more insight into the dynamics of our model, we study the 
two-dimensional equations of motion (\ref{eq:2b}) and (\ref{eq:2c}) 
with the aim to understand properties of 
the stationary probability density $P(v,\eta)$.

\begin{figure}[h]
\includegraphics[width=0.43 \textwidth]{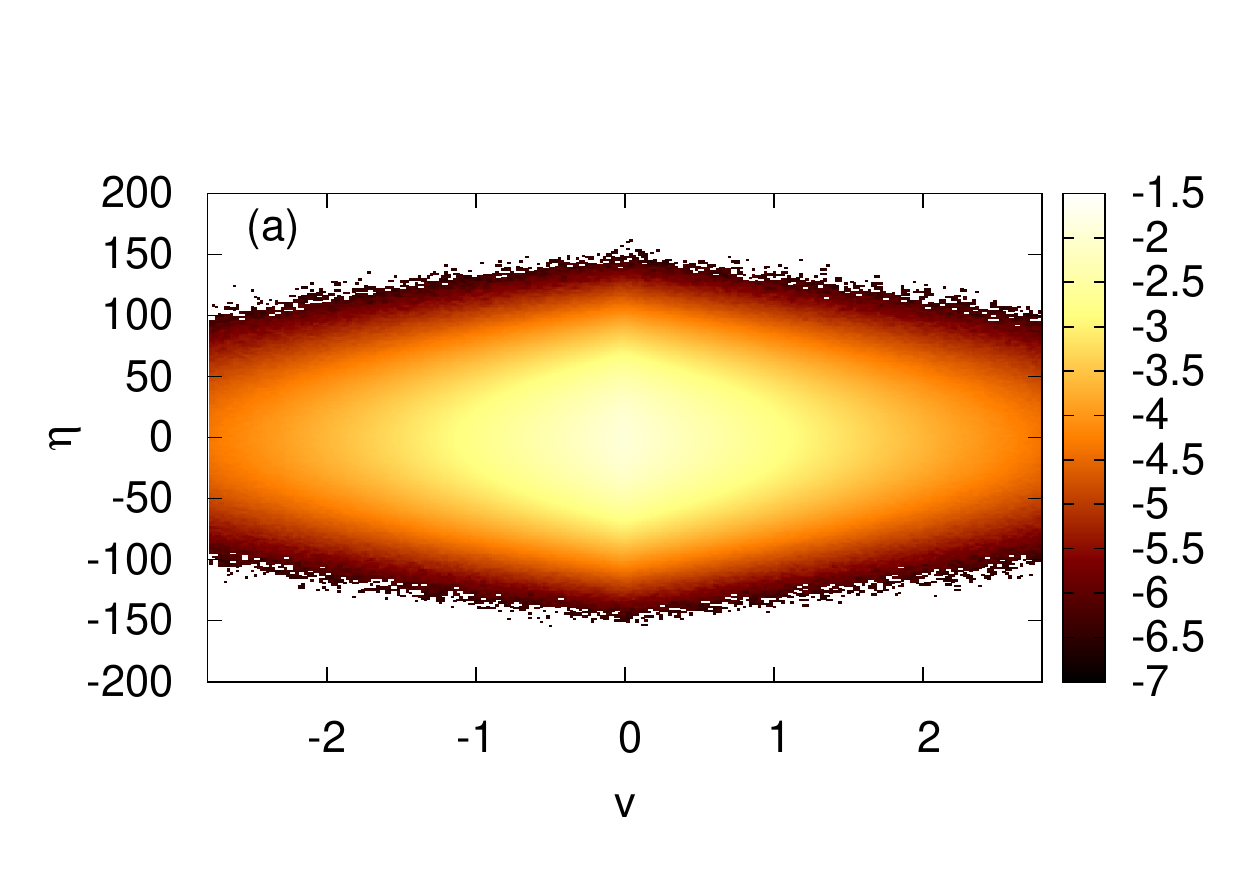}\\
\includegraphics[width=0.43 \textwidth]{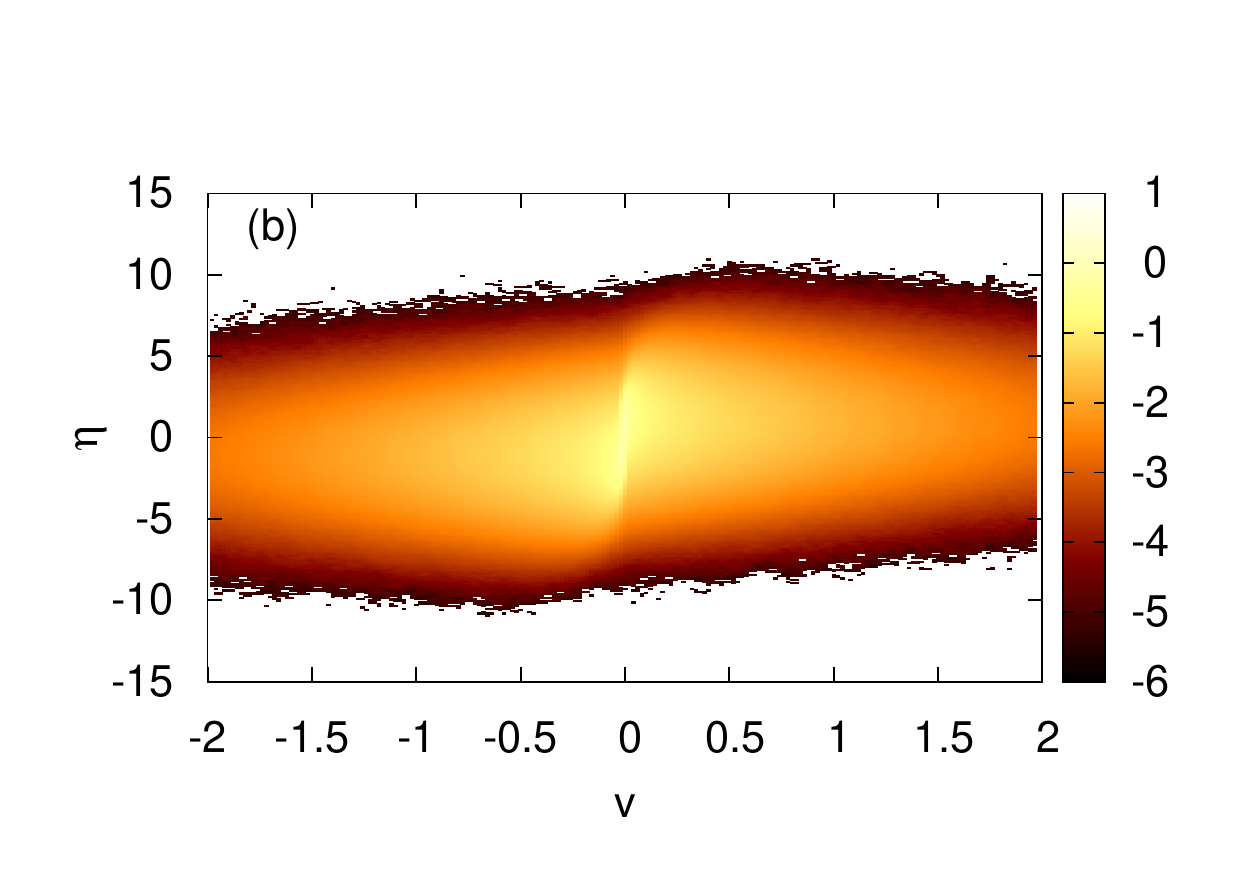}\\
\includegraphics[width=0.43 \textwidth]{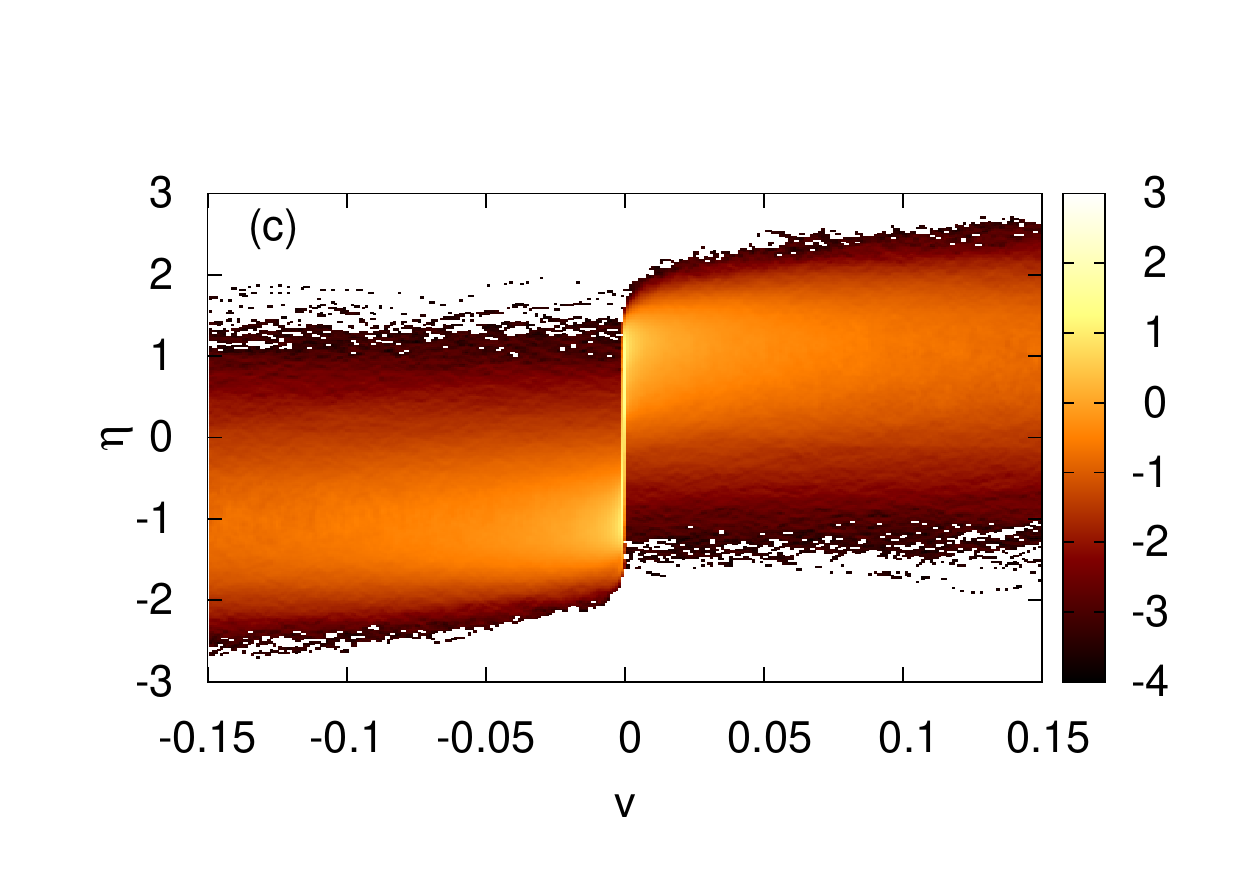}
\caption{Logarithmic density plot of the
stationary probability density, obtained from numerical simulations of 
eqs.(\ref{eq:2b}) and (\ref{eq:2c}), for different values of the
correlation time: $\tau=0.001$ (a), $\tau=0.1$ (b), 
$\tau=1.0$ (c). The density has been
sampled with a resolution of $\Delta v = 0.002$ and 300 bins in $\eta$-direction. Slices of the density at $v=0$ can be found in figure \ref{fig:2g}.}\label{fig:2e}
\end{figure}

To begin with we perform numerical simulations of the dynamics of 
eqs.(\ref{eq:2b}) and (\ref{eq:2c}) (see above for details of the numerical 
integration scheme). Density plots on a logarithmic scale of the full
stationary distribution (regular and singular part) are shown in figure \ref{fig:2e}.
For $\tau = 0.001$ the singular part hardly matters and results are
almost  indistinguishable from the white noise case within the resolution 
of our simulations. The regular
density shows a Gaussian profile in the $\eta$ direction as well as
exponential decay in the $v$ direction.
By increasing $\tau$, the density changes significantly 
as the singular part becomes noticeable (cf. eq.(\ref{eq:2m}) and
figure \ref{fig:2c}). Furthermore, the
regular part of the density becomes asymmetric as the
two components in the half spaces $v>0$ and $v<0$ are shifted 
against each other. 

\begin{figure}[h]
\includegraphics[width=0.43 \textwidth]{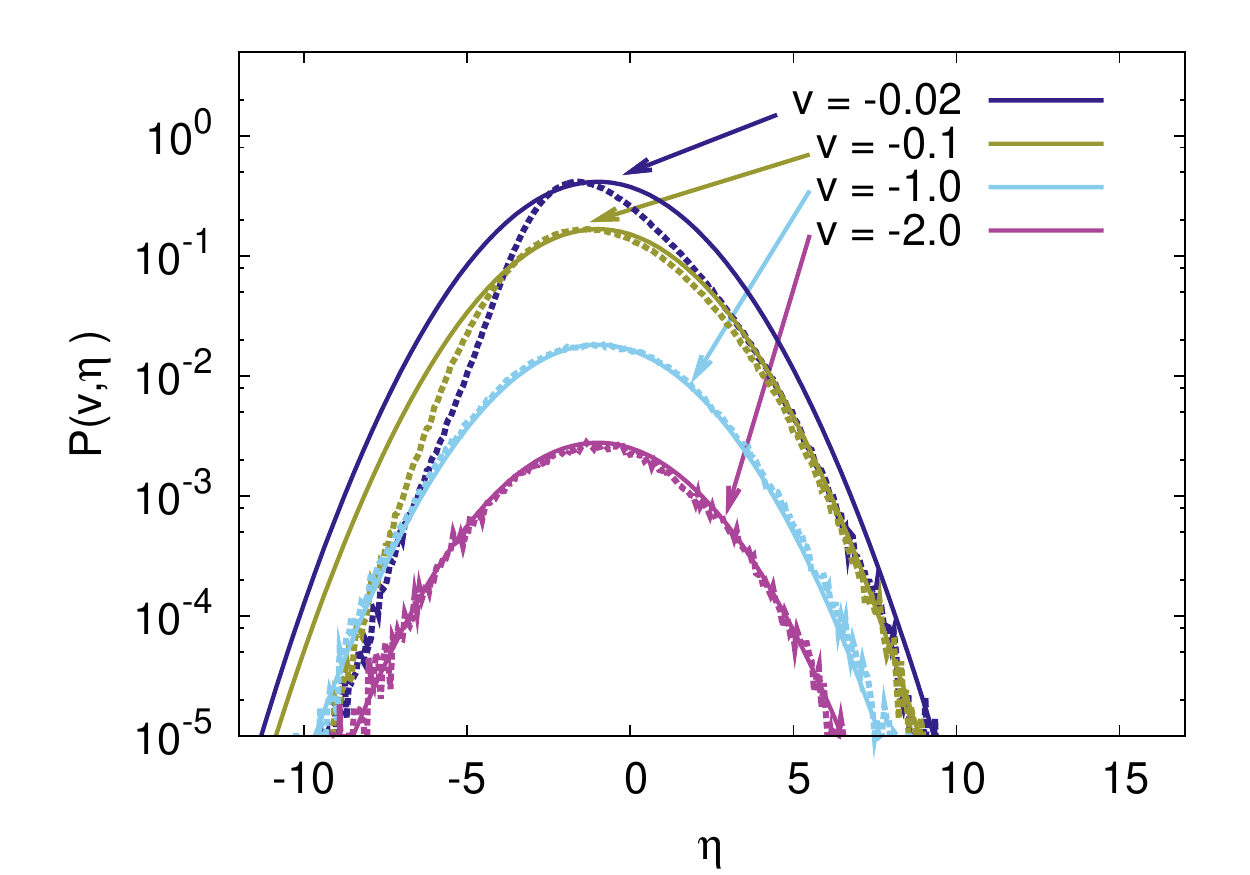}
\caption{Regular component of the stationary probability density at
$\tau=0.1$. Dependence on the noise amplitude for different fixed
values of the velocity $v$. Results of numerical simulations (dashed lines) and the
analytic asymptotic expression (solid lines), eq.(\ref{eq:2o}). The normalisation of the analytics is fitted to the numerical data.} \label{fig:2h}
\end{figure}

\begin{figure}[h]
\includegraphics[width=0.43 \textwidth]{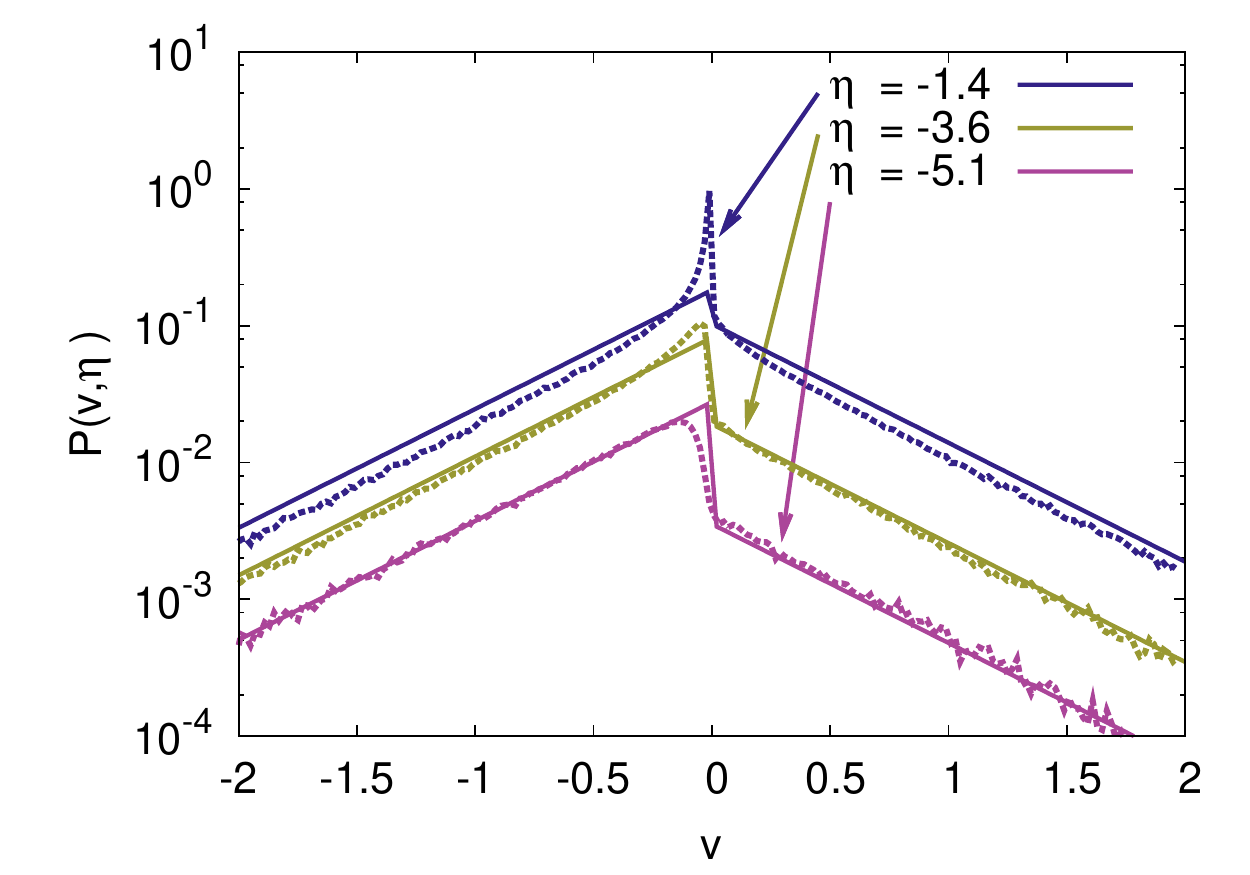}
\caption{Regular component of the stationary probability density at
$\tau=0.1$. Dependence on the velocity for different fixed
values of the noise amplitude $\eta$. Results of numerical simulations 
(dashed lines) and the analytic asymptotic expression (solid lines), eq.(\ref{eq:2o}). The normalisation of the analytics is fitted to the numerical data.} \label{fig:2l}
\end{figure}

For further analytical insight we try to formulate the corresponding
Fokker-Planck system. Using the regularized version of the equations of 
motion, eqs.(\ref{eq:2b}) and (\ref{eq:2d}), the Fokker-Planck equation reads
\begin{equation}\label{eq:2n}
\partial _t P = \partial _v 
\left(\sigma_\varepsilon (v) - \eta \right)P 
+ \partial _\eta \left(\frac{\eta}{\tau} 
+ \frac{1}{2 \tau ^2}\partial _\eta \right)P \, .
\end{equation}
There is no obvious way to compute the stationary solution because
detailed balance is violated. The marginal distribution for
the noise amplitude is however easily computed as
\begin{equation}\label{eq:2q}
P_\eta(\eta) = \sqrt{\frac{\tau}{\pi}} \exp (- \tau \eta ^2 ) 
\end{equation}
and does not depend on the regularisation. Hence eq.(\ref{eq:2q}) applies
as well in the dry friction limit $\varepsilon \rightarrow 0$, which
does not come as a surprise (cf. eq.(\ref{eq:2b})).
In the dry friction limit the expression
\begin{equation}\label{eq:2o}
P(v,\eta) = \exp \left(-2|v| + 2 \tau\sigma_0 (v) \eta - \tau \eta^2 \right)
\end{equation}
formally solves the stationary Fokker-Planck equation, see eq.(\ref{eq:2n})
in the limit $\varepsilon \rightarrow 0$, as long as
$v$ is nonzero. It differs from the regular part of the marginal (eq.(\ref{eq:2m})) by the sign of the mixed $(v, \eta)$ term. Certainly eq.(\ref{eq:2o}) does not provide an
analytic solution for the stationary density
as eq.(\ref{eq:2o}) does not obey the required matching
conditions at $v=0$. Nevertheless, if the impact
of the stick-slip phenomenon at $v=0$ remains localised
then eq.(\ref{eq:2o}) provides the
asymptotic behaviour for large values of velocities. This
assertion can be verified by looking into the numerical data. 
In figure \ref{fig:2h} slices of the regular density taken at
constant values of the velocity show deviations from the Gaussian
profile close to the singular component, i.e., at low velocities.
However, the Gaussian profile according to eq.(\ref{eq:2o}) is restored 
when we increase the velocity, i.e., at regions in phase space further
away from the sticking region. Deviations from the Gaussian profile or strictly speaking the asymmetry of the distribution in $\eta$ direction can also be observed in figure \ref{fig:2e} (bottom) for a high noise correlation $\tau$. A similar feature is displayed by slices
taken at constant noise level, see figure \ref{fig:2l}. 


The singular part of the distribution, that means the dynamics
of sticking particles is entirely governed by
eq.(\ref{eq:2b}), that means by the Fokker-Planck equation of the
Ornstein-Uhlenbeck process. But natural 
boundary conditions do not apply as particles perform stick-slip transitions.
\begin{figure}[h]
\centerline{\includegraphics[width=0.43 \textwidth]{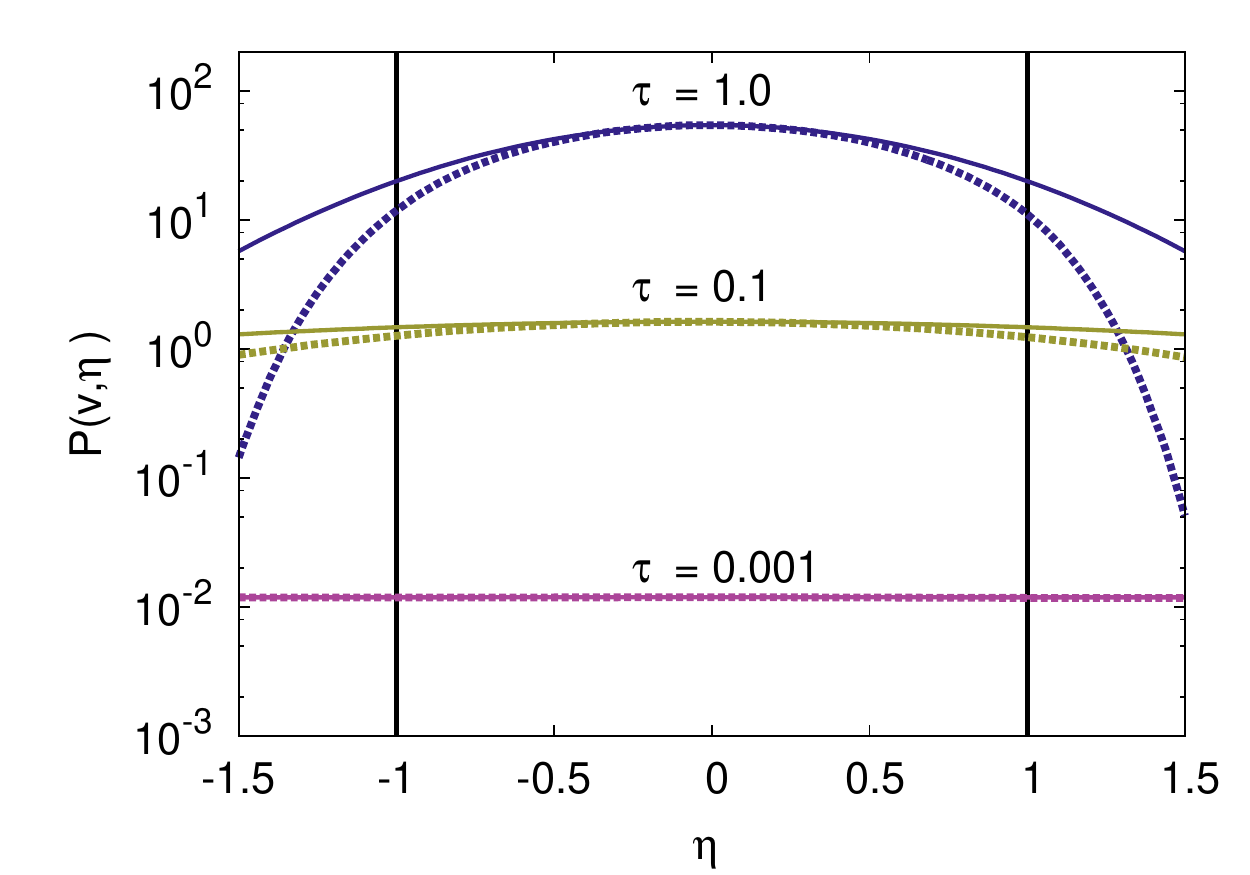}}
\caption{Slices of the stationary density at $v=0$ in $\eta$ direction, obtained
from numerical simulations (dashed lines)
and analytic results for the marginal distribution, 
eq.(\ref{eq:2q}) (solid lines) for different values of the noise correlation time. The vertical solid lines indicate the region of the singular part of the distribution (sticking regime). The normalisation of the analytics is fitted to the numerical data.}\label{fig:2g}
\end{figure}
For the singular part of the density at $v=0$ we have already indicated that
increasing the correlation time results in a considerable decrease of
the probability of sliding particles. As a result the main contribution to the
marginal distribution, eq.(\ref{eq:2q}), will come from the density at $v=0$
as, apart from an exponentially small contribution, particles become
immobile. That is in quantitative agreement with direct
simulations, see figure \ref{fig:2g}. For small values of the correlation
time considerable deviations from the normal distribution appear as 
on the one hand particles become mobile frequently and on the
other hand there is a constant stream of mobile particles getting
stuck (see as well figures \ref{fig:2a} - \ref{fig:2b1}).

\begin{figure}[h]
\centerline{\includegraphics[width=0.43 \textwidth]{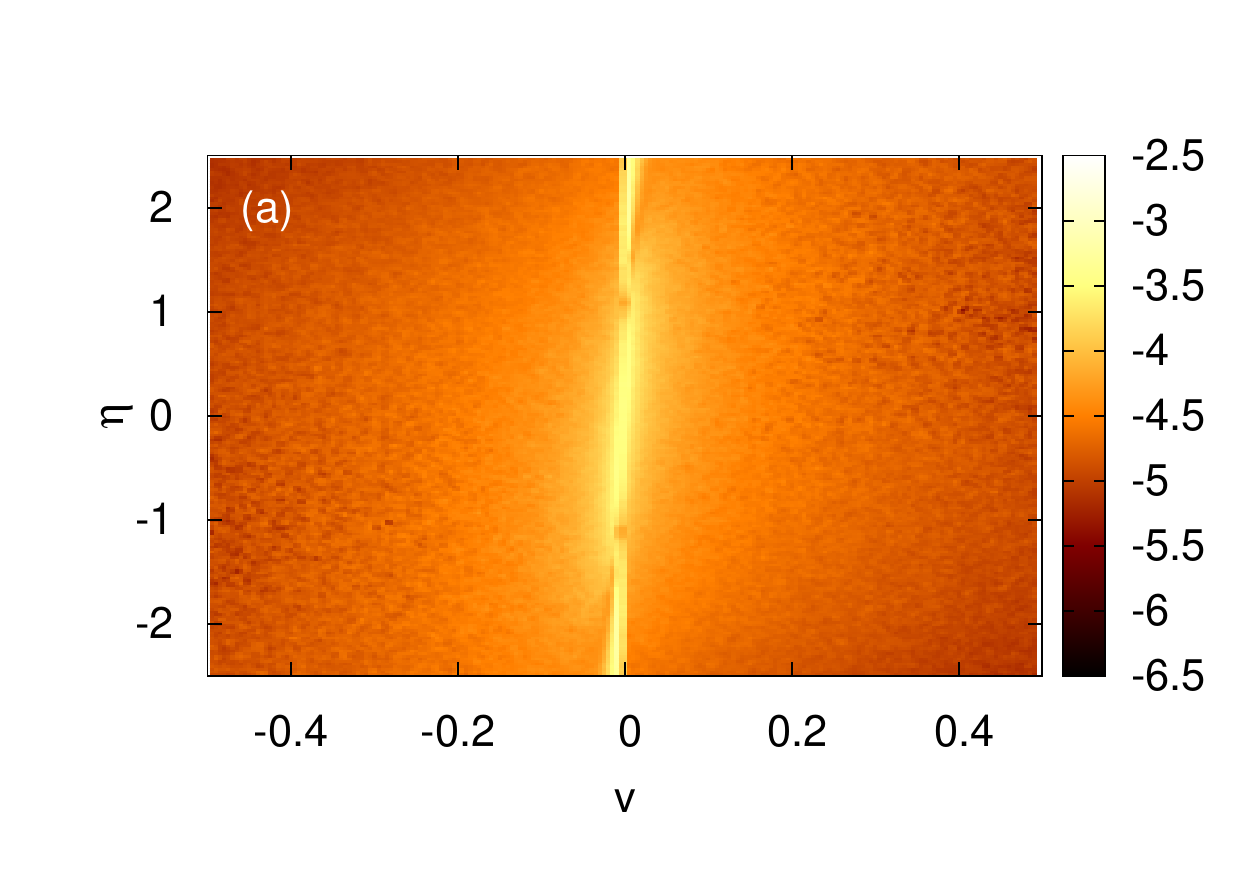}}
\centerline{\includegraphics[width=0.31 \textwidth]{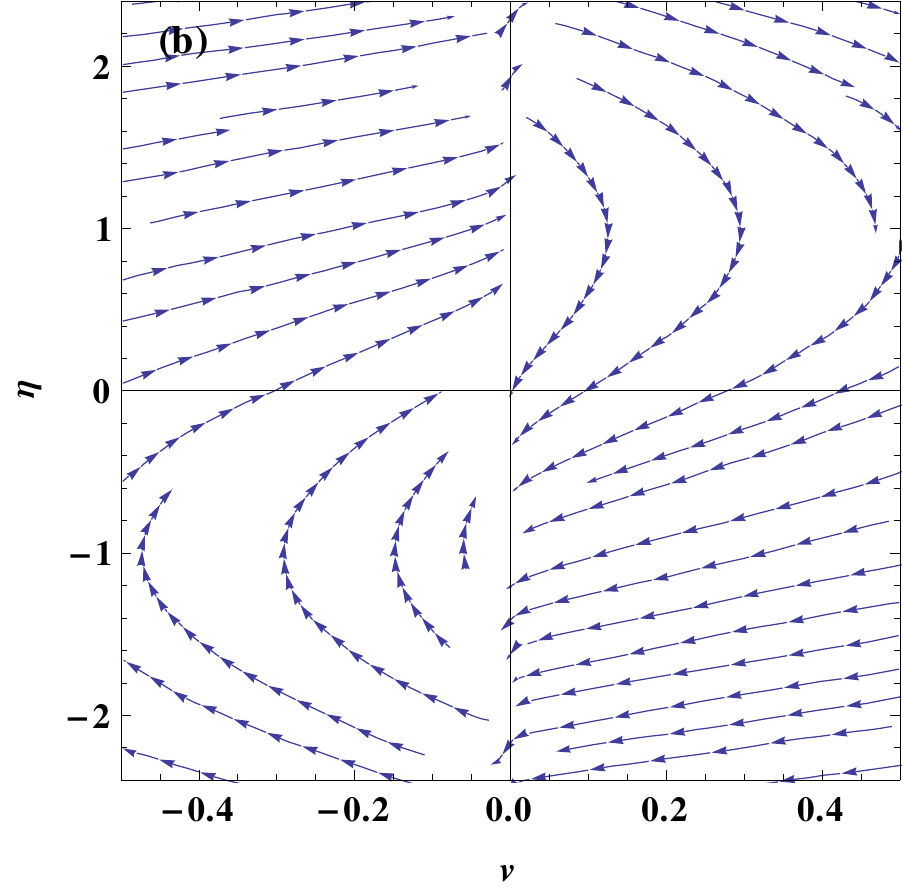}}
\caption{Logarithmic density plot (a) and stream plot (b) of the regular part of the stationary 
probability current of the system eqs.(\ref{eq:2b}) and (\ref{eq:2c})
for $\tau = 0.1$, obtained from numerical simulations. The density plot shows the absolute value of the current in the $(v,\eta)$ plane, whereas the stream plot displays the normalised vector field of the current.}\label{fig:2i}
\end{figure}

\begin{figure}[h]
\centerline{\includegraphics[width=0.43 \textwidth]{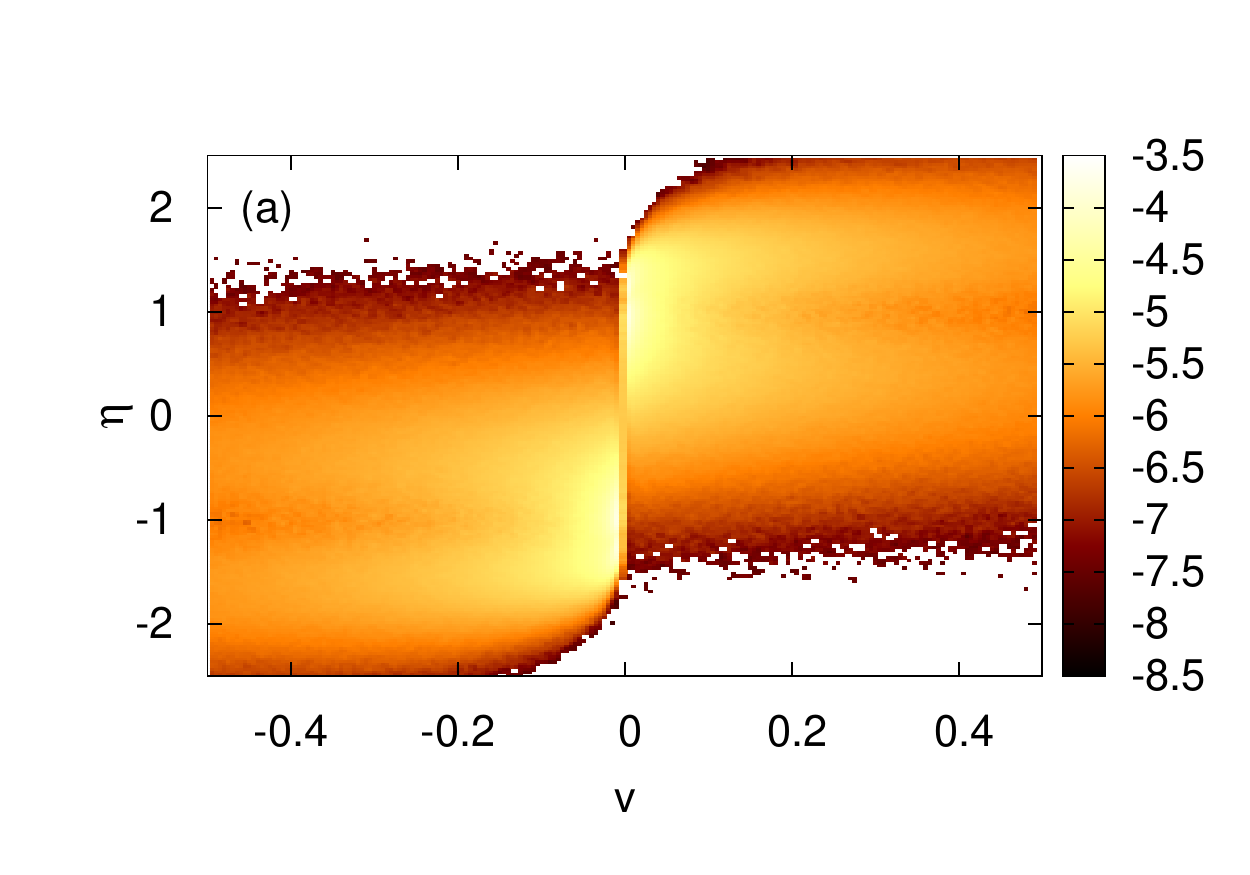}}
\centerline{\includegraphics[width=0.31 \textwidth]{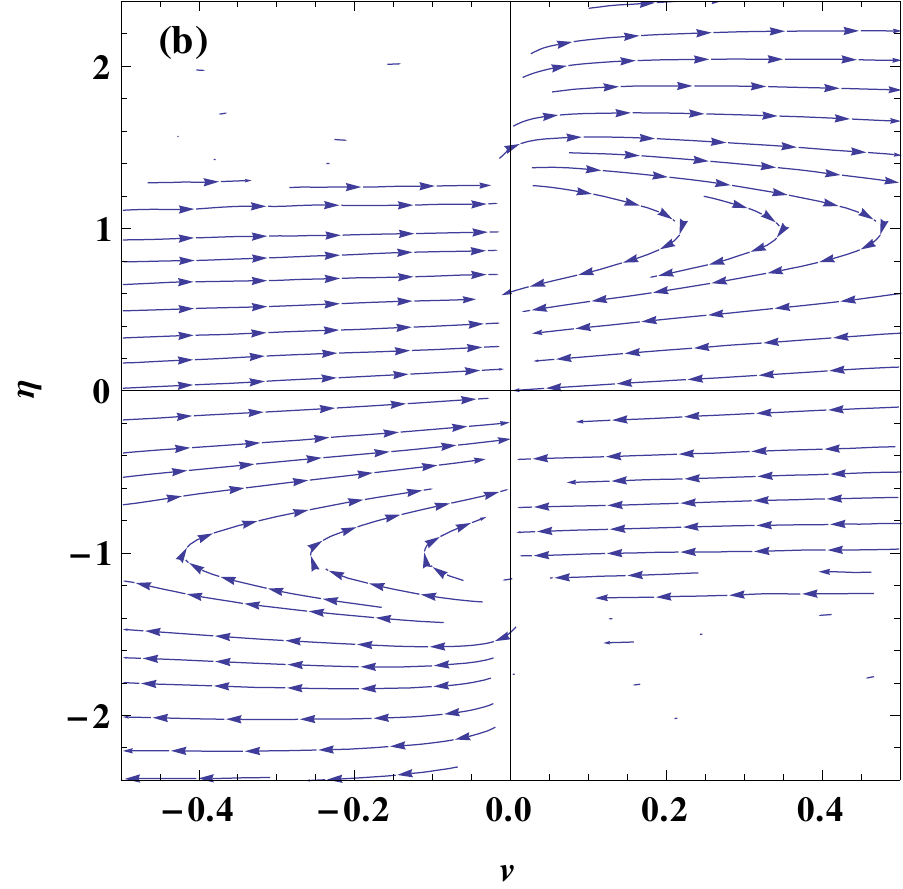}}
\caption{Logarithmic density plot (a) and stream plot (b) of the regular part of the stationary 
probability current of the system eqs.(\ref{eq:2b}) and (\ref{eq:2c})
for $\tau = 1.0$, obtained from numerical simulations. The density plot shows the absolute value of the current in the $(v,\eta)$ plane, whereas the stream plot displays the normalised vector field of the current.}\label{fig:2j}
\end{figure}

\begin{figure}[h!]
\includegraphics[width=0.43 \textwidth]{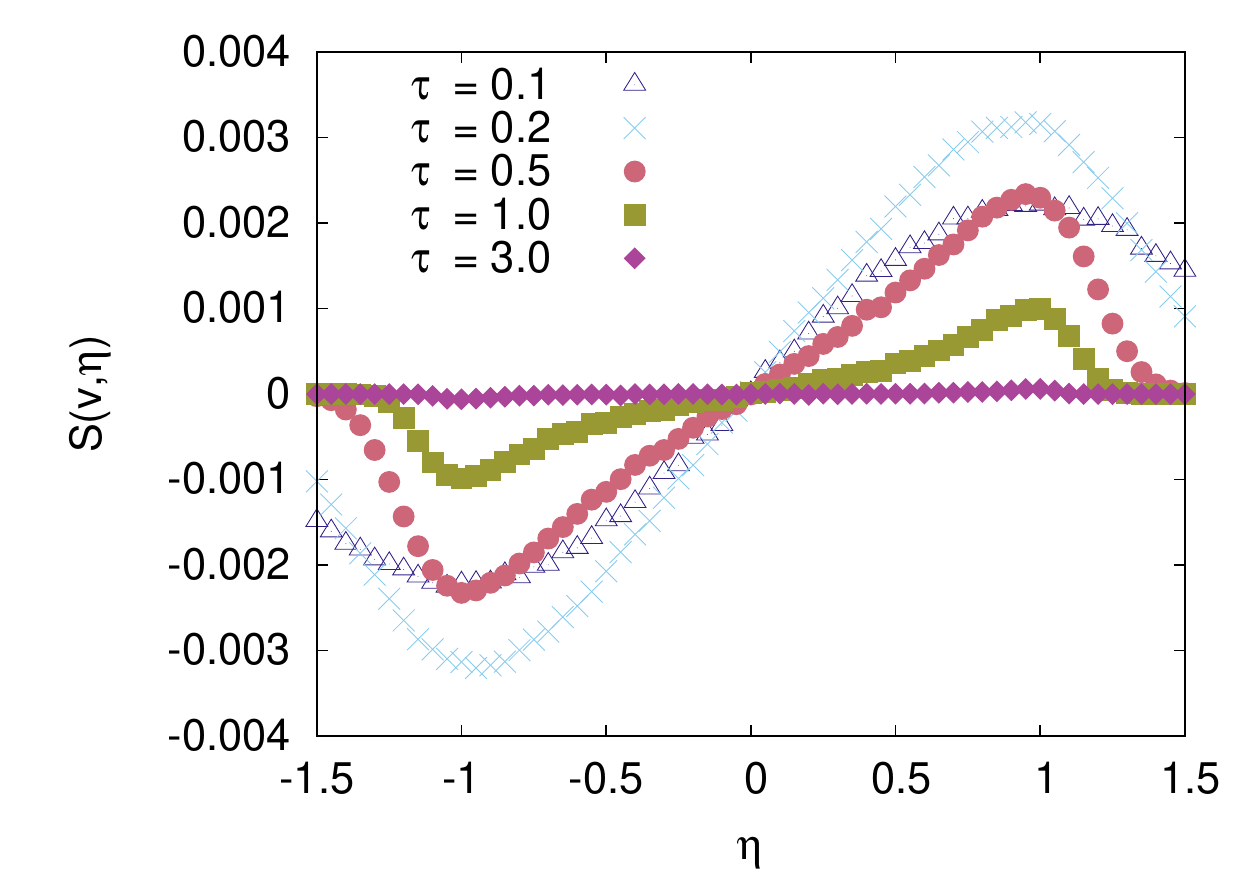}
\caption{$\eta$ component of the stationary probability current at $v=0$ 
for different values of the $\tau$, obtained from numerical simulations of eqs.(\ref{eq:2b}) and (\ref{eq:2c}).}\label{fig:2k}
\end{figure}

In the region of the phase space close to the stick-slip transition
the shape of the distribution is affected at intermediate values of the noise correlation. There is
no obvious way to tackle the issue by analytical means, e.g.,  
with the matching conditions between the singular and the regular part. But
one can at least have a closer look at the probability current which is
a clear signature of the nonequilibrium properties of our model.
By using the method from \cite{JUS03} we compute the 
probability current directly from the time series of our model (eqs.(\ref{eq:2b}), (\ref{eq:2c})) for different values of $\tau$, see figures \ref{fig:2i} and \ref{fig:2j}.
The entire flow pattern is symmetric, and the main part of the current
is concentrated in regions with low velocity $v$. For $\eta>1$ the
current predominantly points in the positive $v$-direction as particles 
are dragged by the external forcing. For larger positive values of $v$ the 
current turns and finally approaches the sticking manifold $v=0$, $|\eta|<1$,
where particles change from sliding to sticking mode.
As the stationary probability current is, by definition, 
solenoidal (see eq.(\ref{eq:2n})) the current on the sticking manifold becomes large and points in the positive or negative direction, see figure \ref{fig:2k}. 
When reaching the critical value $|\eta|=1$ particles start sliding again.
In particular, the current on the sticking 
manifold $v=0$ and $|\eta|<1$ and the current entering or leaving
this manifold obey matching conditions.
Hence, for the piecewise-smooth dynamics one can write
down a system of two coupled Fokker-Planck equations, one governing the 
sticking and one governing the sliding motion, with appropriate matching 
conditions and source terms. It is however not obvious whether such a 
formulation for the regular and the singular component of the probability
distribution would give more insight than direct numerical 
simulations of the associated Langevin equation, let along providing 
a pathway for an analytic approach.

Figure \ref{fig:2k} indicates a non-monotonic behaviour of the current by varying the correlation time of the noise $\tau$. For low values of $\tau$, the current is very small, as we are close to the white noise limit. By increasing $\tau$ the current increases as well up to a value close to $\tau = 0.2$. Increasing $\tau$ further the current decreases and almost vanishes, see the results for $\tau = 3.0$ in figure \ref{fig:2k}. For higher values of the noise correlation the probability current decays rapidly outside of the interval $\eta \in (-1,1)$. In view of the particular structure of the stationary density this is
hardly surprising, as the singular component of the density dominates for
high correlation times, the dynamics is dominated by sticking
particles, and only a small part of the probability density contributes to the
sliding motion and finally to the probability current.

\section{\label{sec:4}Dynamical properties of the piecewise linear model}

Traditional correlation functions are a useful tool to study
dynamical properties, in particular, within the context of
linear response theories. From a theoretical perspective
their analytical properties are related with the 
eigenvalue spectrum of the underlying equations of motion, 
e.g. with the spectrum of Fokker-Planck operators. Furthermore
correlation functions are experimentally accessible and they allow
to introduce the concept of correlation times. As a shortcoming
correlation functions may not allow for the proper characterisation
of intermittent behaviour, such as stick-slip transitions, which
needs then to be addressed separately by a suitable statistical
measure.

\subsection{\label{sec:4a}Power spectral density}

To begin with, we want to investigate how the correlation time of the 
noise $\tau$ influences  the correlations in our system. To be slightly
more precise we will discuss the $\tau$-dependence of the power 
spectral density of the velocity $v$, and  the corresponding
linewidth. The latter provides 
insight into the structure of the eigenvalue spectrum of an underlying 
Fokker Planck operator governing the dynamics of the system. For the dry 
friction model with white noise, $(\tau = 0)$, a spectral gap between 
the two first eigenvalues has been observed \cite{RIS96,TOU10}. 
In \cite{TOU12} a closed expression for the power spectral density of the velocity has been derived based on the Laplace transform of the propagator. 

For noise with finite correlation time
we mainly rely on numerical investigations 
since analytic expression for the stationary probability density are
unknown. We calculate the power spectral density of the variable 
$v$ by averaging over 800 numerically generated time traces of length
$T=10^{4}$. We base our analysis on the autocorrelations 
of the velocity. Hence, the corresponding power spectral density predominantly 
probes properties of the sliding phase as velocities vanish in the 
sticking phase.

\begin{figure}[h!]
\center
\includegraphics[width=0.43 \textwidth]{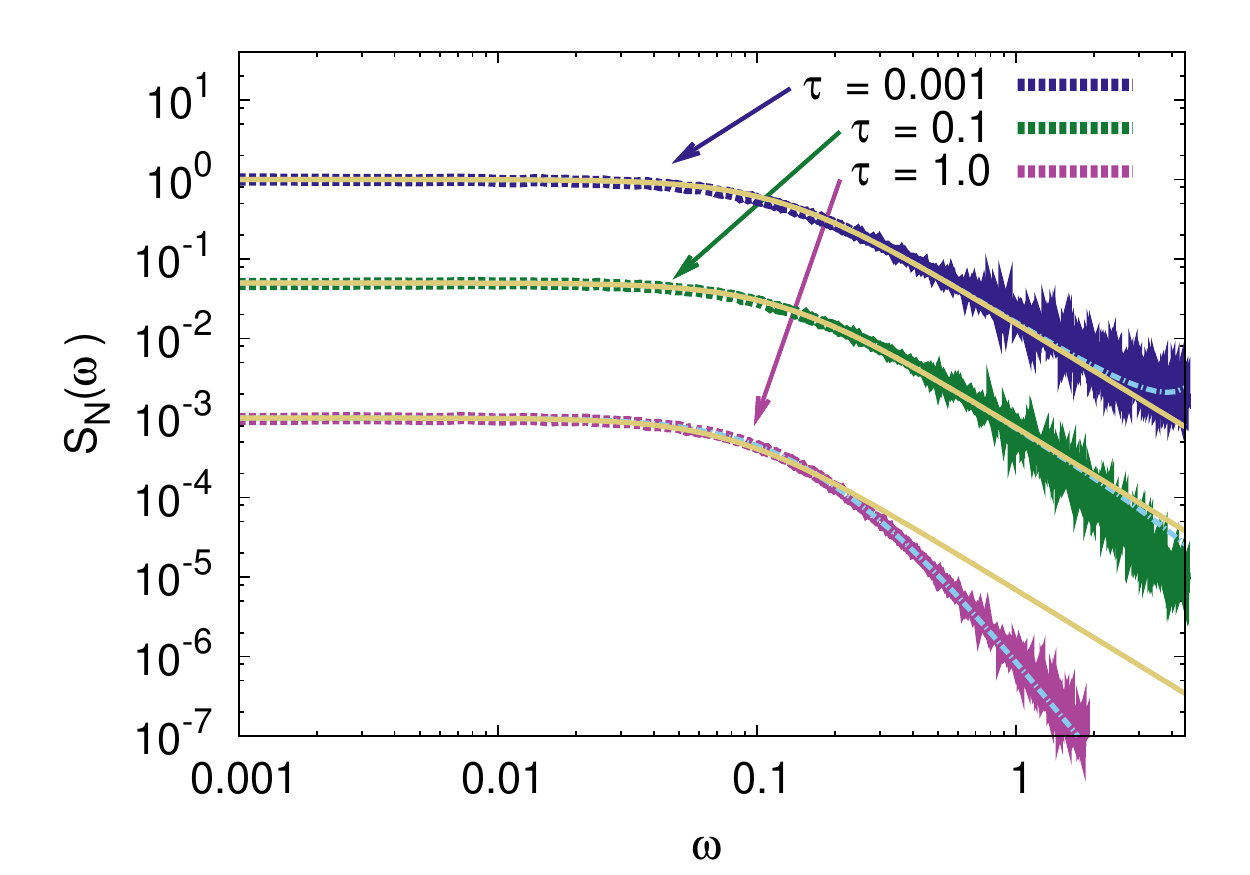}
\caption{Double-logarithmic plot of the power spectral density of the 
$v$-variable for $\tau = 0.001,~0.1,~1.0$ (from top to bottom) and different fit functions. Spectra have been normalised
by the condition $S_N(0)=1$ and shifted respectively. Numerical simulations (dashed lines), Lorentzian fit ($\sim 1/(1 + \omega ^2)$) ((bronze) solid lines), quartic spectral fit ($\sim 1/(1 + a \omega ^2 + b \omega ^4)$) ((cyan) dot-dashed lines).}
\label{fig:3b}
\end{figure}

Figure \ref{fig:3b} shows the numerical results of the normalised spectral 
densities for different values of $\tau$. The normalised power spectral densities 
$S_N(\omega)$ have a single central peak at $\omega=0$
indicating an exponential decay of the corresponding autocorrelation function.
For small values of $\tau$, and in accordance with the white noise limit,
$S_N(\omega)$ is a Lorentzian
with power law behaviour $\omega^{-2}$ at an intermediate frequency range.
Such decay changes when increasing the noise correlation time $\tau$,
resulting in a decay proportional to $\omega^{-4}$ at medium frequencies.
The corresponding analytic behaviour indicates a smooth
autocorrelation function at time zero.

\begin{figure}[h!]
\centerline{\includegraphics[width=0.43 \textwidth]{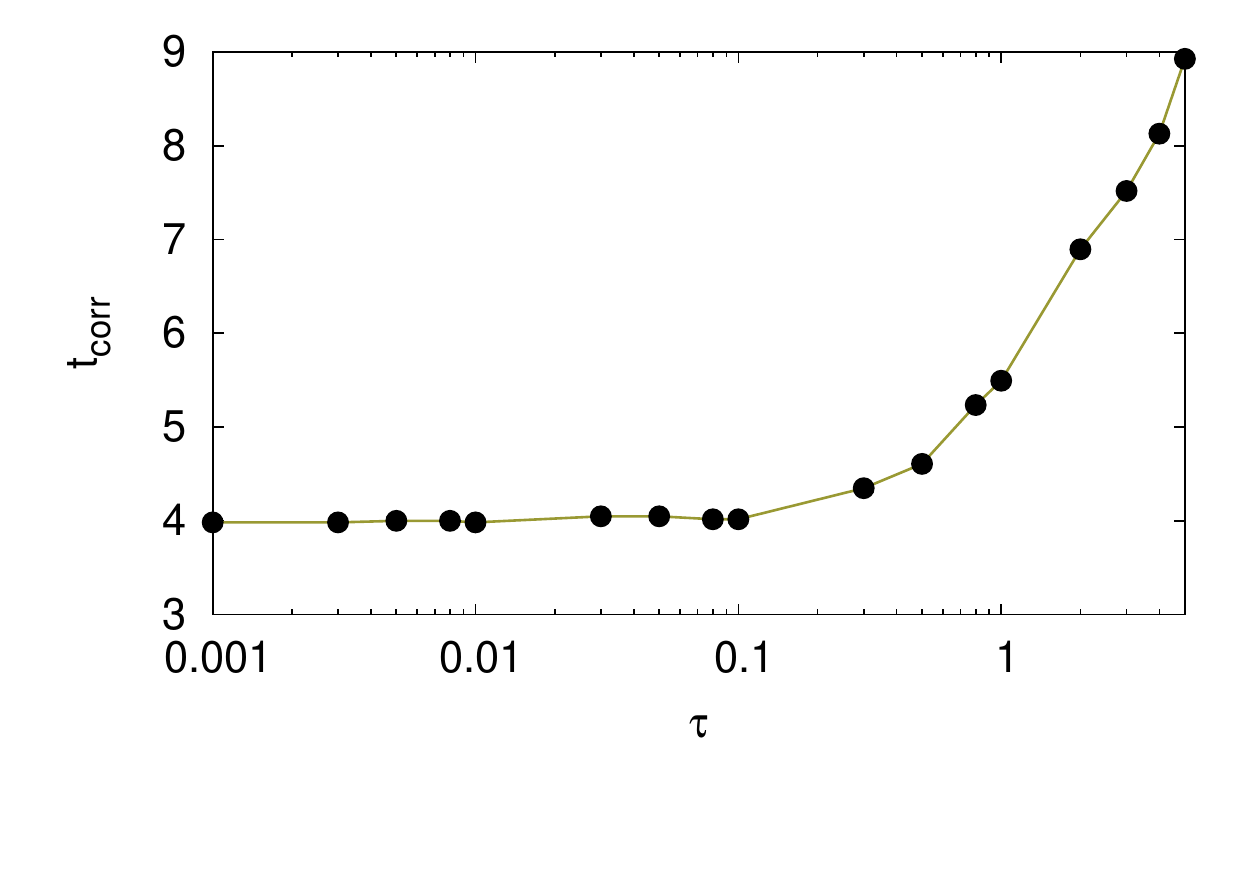}}
\caption{Correlation time $t_{corr}$ as a function of the
noise correlation time $\tau$, obtained from numerical simulations of the spectral density and estimating the full width at half maximum by using a quartic spectral fit, see figure \ref{fig:3b}.}\label{fig:3c}
\end{figure}

The complex valued singularities of the power spectral
density are signatures of the the non-vanishing eigenvalue of 
an underlying Fokker Planck operator.
For power spectral densities with a well defined central peak,
the full width at half maximum $\Delta \omega$ can be 
related to the correlation time of the system $t_{corr}$ via the 
Wiener-Khinchin theorem. Following results for linear stochastic processes
we define here a correlation time by $t_{corr}=1/\Delta \omega$. 
Using a fit function of the form $1/(1 + a \omega ^2 + b \omega ^4)$ 
for the power spectral densities $S_N(\omega)$
we evaluate the correlation time, see figure \ref{fig:3c}.

The correlation time $t_{corr}$ essentially coincides with the value
of the white noise limit as long as $\tau<0.1$. While there is no
sharp transition, $t_{corr}$ significantly increases monotonically
when the noise correlation time exceeds a ``critical'' value of $\tau=0.1$.
Hence, signatures of the stick-slip transition become dynamically 
visible above such a critical value. The transition-like feature is
in accordance with the findings about the stationary
density reported in the previous section, e.g. see figure \ref{fig:2d}.

\subsection{\label{sec:4b}Distribution of sticking and sliding periods}
The time traces shown in figures \ref{fig:2a} - \ref{fig:2b1} suggest a closer relation
of the dynamics with intermittency phenomena. To probe directly the 
dynamical features of the stick-slip transition
we look at the distribution of sticking and sliding times, i.e.,
the distribution of time intervals the particle spends in states $v=0$ and 
$v\neq 0$. 

\begin{figure}[h!]
\centerline{\includegraphics[width=0.43 \textwidth]{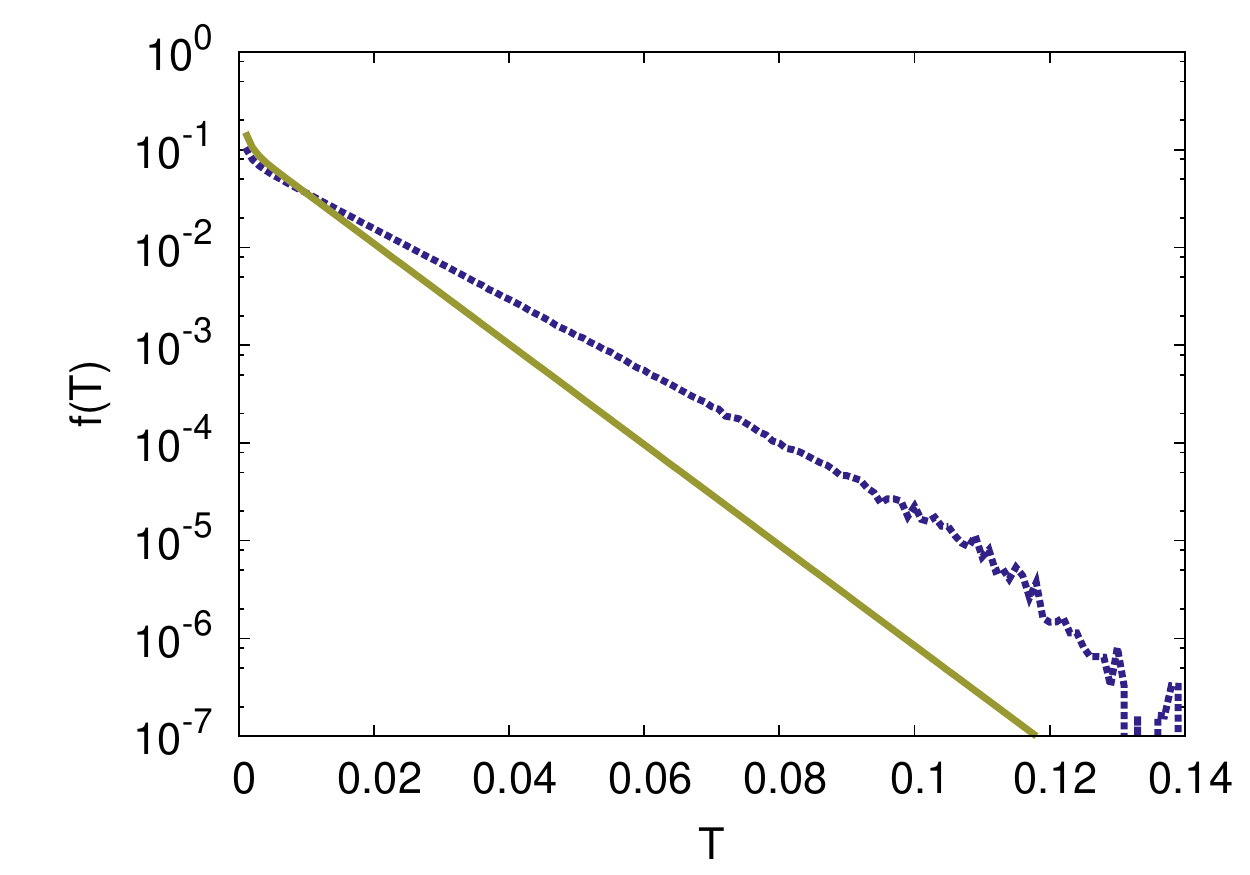}}
\caption{Distribution of sticking time intervals, $f(T)$, 
on a semi-logarithmic scale for $\tau=0.1$, obtained numerically ((blue) dashed line) and semi-analytically from the exit time problem for the Ornstein-Uhlenbeck process ((bronze) solid line) (the inverse Laplace Transform of eq.(\ref{eq:2r})).} \label{fig:4a}
\end{figure}

\begin{figure}[h!]
\centerline{\includegraphics[width=0.43 \textwidth]{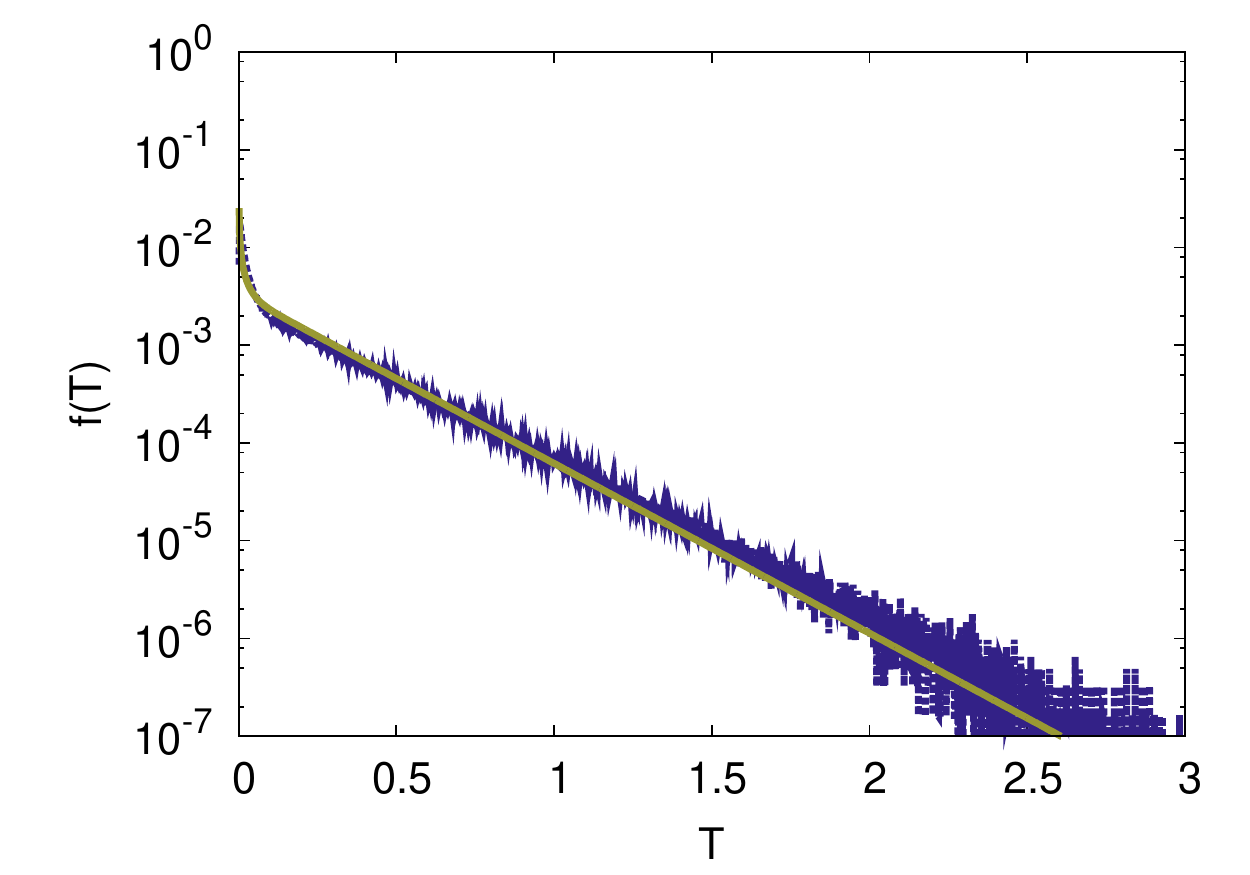}}
\caption{Distribution of sticking time intervals, $f(T)$, 
on a semi-logarithmic scale for $\tau=0.5$, obtained numerically ((blue) dashed line) and semi-analytically from the exit time problem for the Ornstein-Uhlenbeck process ((bronze) solid line) (the inverse Laplace Transform of eq.(\ref{eq:2r})).} \label{fig:4b}
\end{figure}

\begin{figure}[h!]
\centerline{\includegraphics[width=0.43 \textwidth]{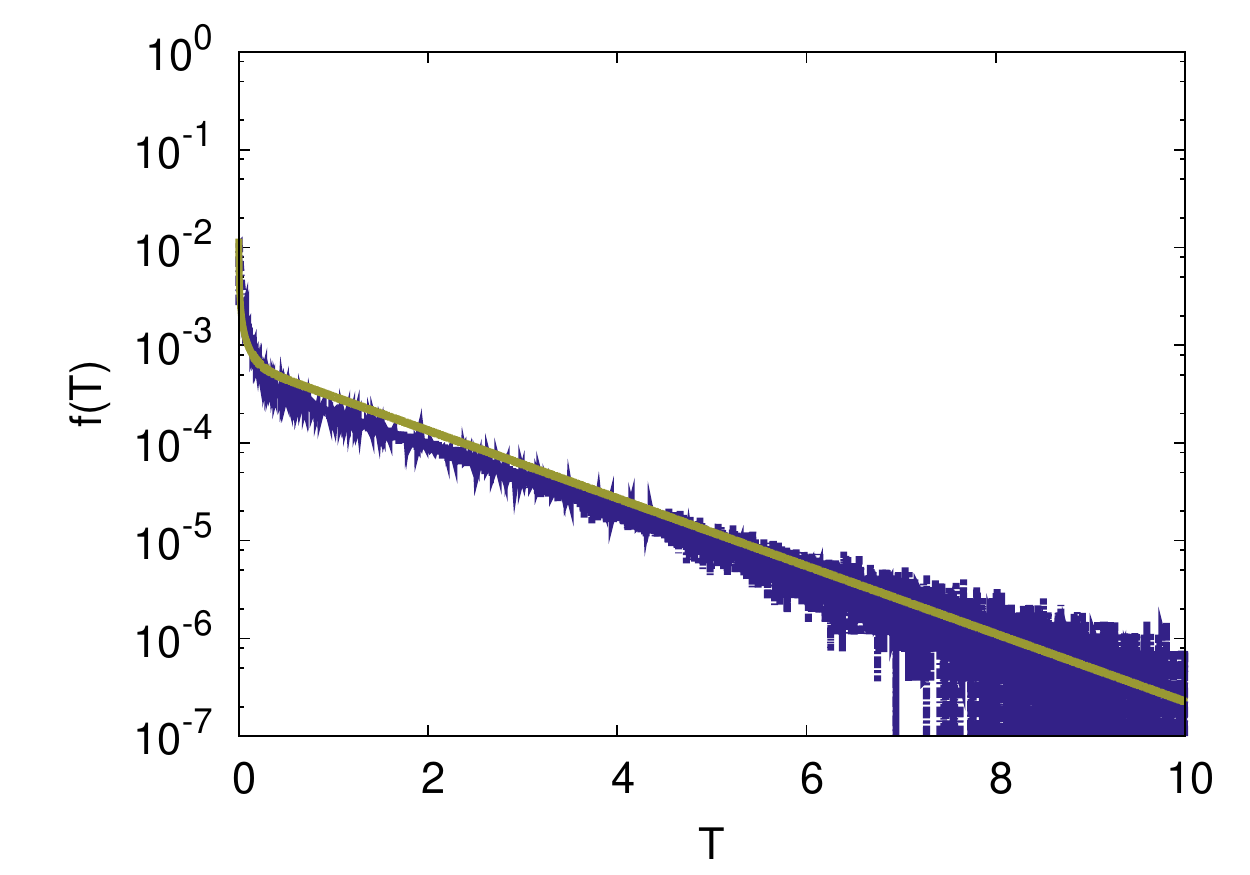}}
\caption{Distribution of sticking time intervals, $f(T)$, 
on a semi-logarithmic scale for $\tau=1.0$, obtained numerically ((blue) dashed line) and semi-analytically from the exit time problem for the Ornstein-Uhlenbeck process ((bronze) solid line) (the inverse Laplace Transform of eq.(\ref{eq:2r})).} \label{fig:4c}
\end{figure}

We start our investigations with the analysis of the sticking time events.
As the dynamics of sticking particles is mainly determined by the exit
time problem of the Ornstein-Uhlenbeck process, see eq.(\ref{eq:2b}), this problem can be treated by analytical means, see \cite{DAH53}.
The Laplace transform of the exit time probability density for an Ornstein-Uhlenbeck process like eq.(\ref{eq:2b}) with symmetric absorbing boundaries ($-a$ and $a$) and a fixed initial condition $|\eta _0| < a$ reads
\begin{eqnarray}\label{eq:2p}
\tilde{f}(s|\eta _0) =& \frac{D_{-s \tau}\left(\sqrt{2\tau} \eta _0 \right) + D_{-s \tau}\left(-\sqrt{2\tau}\eta _0 \right)}{D_{-s \tau}\left(\sqrt{2\tau}a\right) + D_{-s \tau}\left(-\sqrt{2\tau}a\right)}\exp \left(\frac{\tau}{2}(\eta _0^2 - a)\right) \nonumber \\
=& \exp \left(\frac{a \tau (a-1)}{2} \right)\frac{\, _1F_1\left(\frac{s \tau }{2};\frac{1}{2};\eta _0^2\tau \right)}{\, _1F_1\left(\frac{ s\tau }{2};\frac{1}{2};a^2\tau \right)},
\end{eqnarray}
where $D_{\nu}(x)$ is the parabolic cylinder function, $\, _1F_1\left(a;b;z \right)$ denotes Kummer's confluent hypergeometric function and we have used some identities for these functions \cite{ABR70}.
We set $a=1$ as the regime, where particles are sticking, is the interval $(-1,1)$, integrate over all possible initial conditions $\eta _0$ within this regime assuming a uniform distribution, to obtain
\begin{eqnarray}\label{eq:2r}
\tilde{f}(s) &=& \frac{1}{2}\int _{-1}^{1}\tilde{f}(s|\eta _0) d\eta _0 \nonumber \\
&=& \frac{\, _1F_1\left(\frac{s \tau }{2};\frac{3}{2};\tau \right)}{\, _1F_1\left(\frac{ s\tau }{2};\frac{1}{2};\tau \right)}. 
\end{eqnarray}
As it is not possible to derive an analytic result for the inverse Laplace Transform of this expression, we use the Talbot method to calculate the exit time distribution numerically \cite{TAL79,ABA04}. The results for certain values of $\tau$ are shown in figures \ref{fig:4a} - \ref{fig:4c}. 
One observes a localised peak in the distribution at $T=0$, and for moderate to large times a simple exponential decay. 
For higher noise correlation times the exponential decay of the distribution becomes smaller. It becomes more likely for particles to stick at $v=0$ which is in accordance with the results in the previous sections.
Our numerical findings for the exit time distribution agree very well with the analytical estimate, i.e. the inverse Laplace transform of eq.(\ref{eq:2r}). It works particularly well for large values of $\tau$ and fails to be valid if we approach the transition value $\tau=0.1$ as stick-slip phenomena become noticeable around this value.

\begin{figure}[h!]
\centerline{\includegraphics[width=0.43 \textwidth]{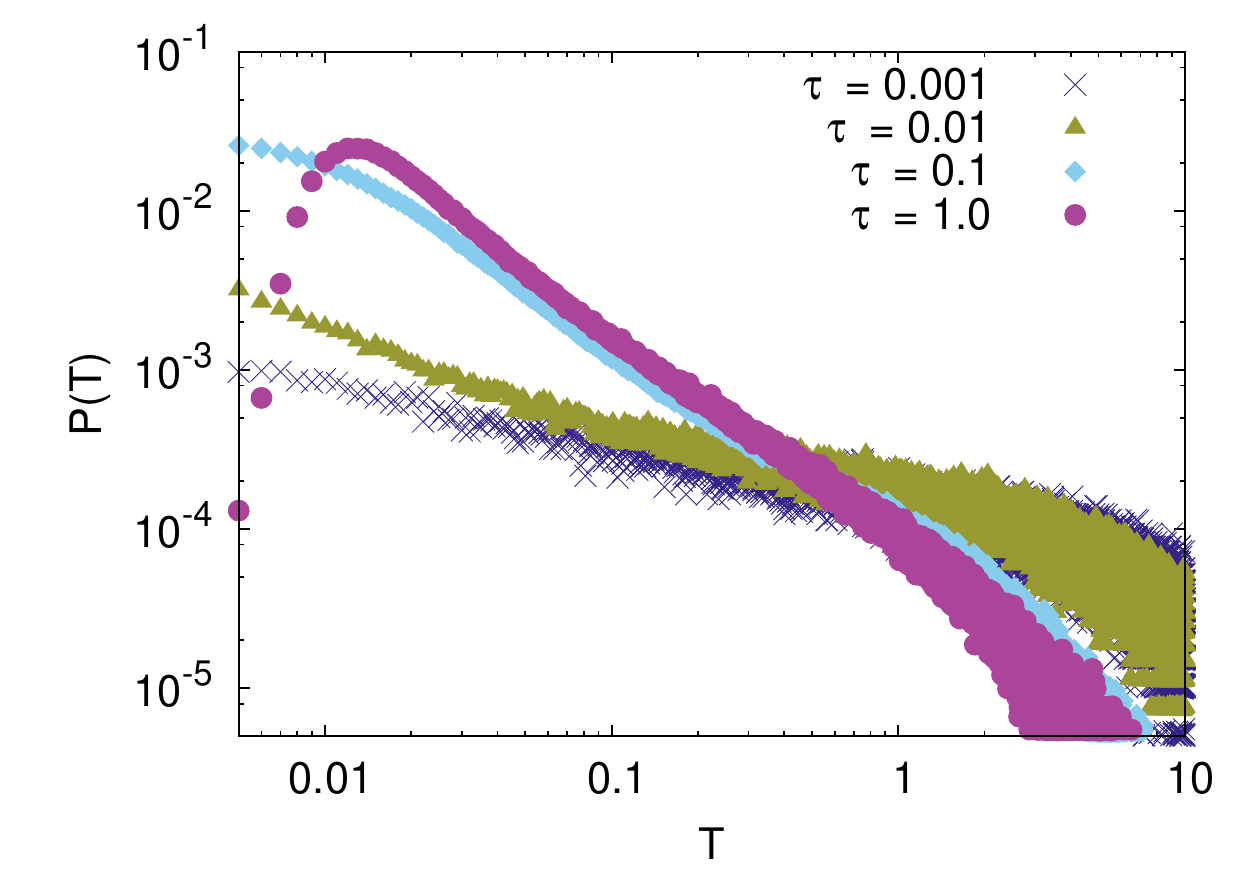}}
\caption{Distribution of sliding time intervals, $P(T)$, 
on a double-logarithmic scale for different values of the noise correlation time, obtained from numerical simulations.} \label{fig:4e}
\end{figure}

\begin{figure}[h!]
\centerline{\includegraphics[width=0.43 \textwidth]{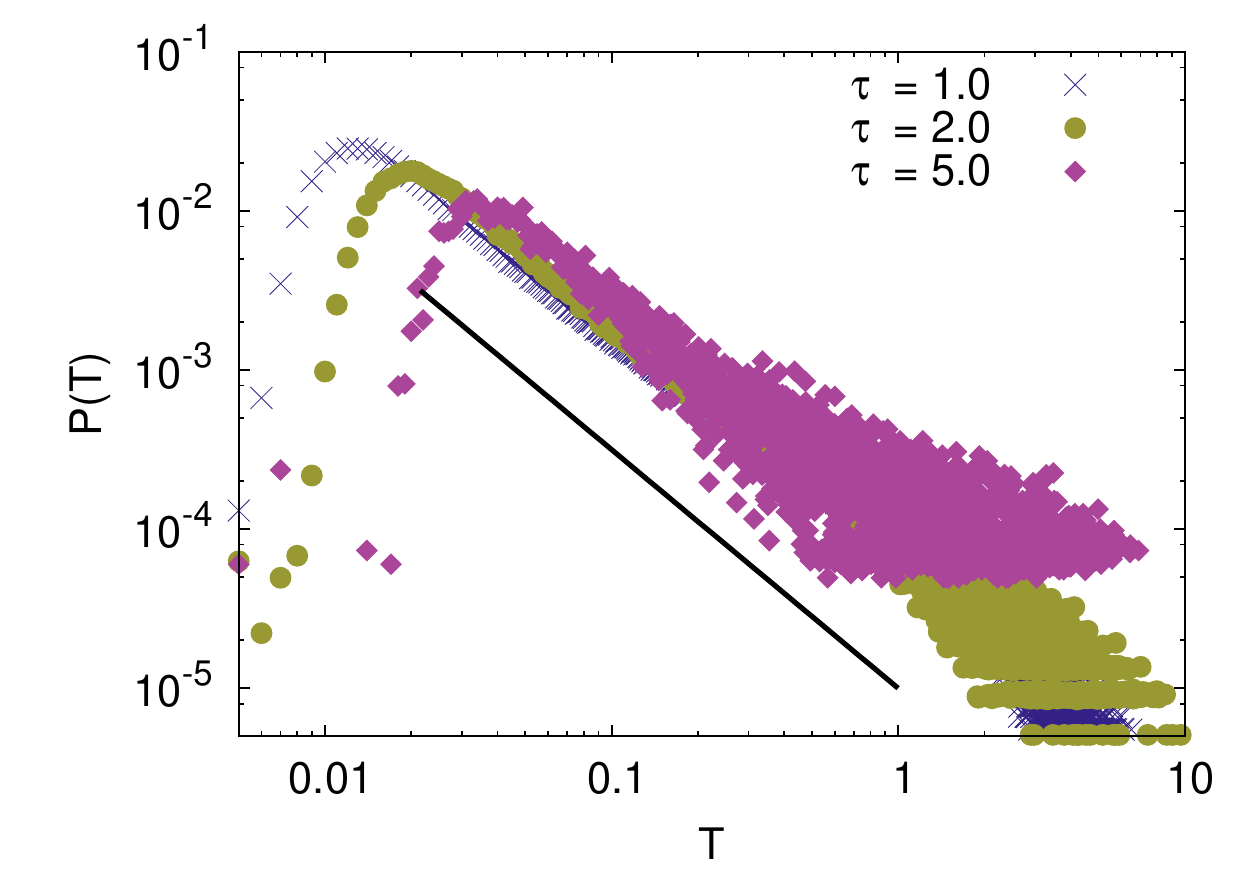}}
\caption{Distribution of sliding time intervals, $P(T)$, 
on a double-logarithmic scale for different values of the noise correlation time, obtained from numerical simulations. The black line
shows a decay according to a power law $T^{-3/2}$.} \label{fig:4f}
\end{figure}

For the remainder of this section 
we focus on the statistics of the sliding events. Figures \ref{fig:4e} and \ref{fig:4f} show the numerically obtained distributions over a wide range
of noise correlation times. For small noise correlation distributions are unimodal with a power law decay at an intermediate range. For a larger noise correlation the distributions develop a maximum at a finite time - so that the most probable sliding time becomes finite.
Figure \ref{fig:4f} indicates a kind of universal behaviour of the distributions at long sliding times $T$ for large correlation times $\tau$.
The asymptotic behaviour of the distributions shows a stunning similarity to characteristics of on-off intermittency \cite{HEA94} as
the power law decay is of the form $T^{-3/2}$ for $\tau \geq 1.0$. But in the context of our model the role of the ``on'' and of the ``off'' state are interchanged as this power law occurs for sliding events.
Looking for an analytic approach the sliding events could be modelled by an exit time problem with constant drift and coloured noise.
For Gaussian white noise and constant drift this problem can be solved analytically \cite{MAJ02}, and the exit time distribution shows an asymptotic behaviour
$P(T,v_0) \sim T^{-3/2}\exp \left(-(T-v_0)^2/2T \right)$. 


\section{\label{sec:5}Conclusion}
We investigated a dry friction model subjected to coloured noise with 
the emphasis on nonequilibrium properties in a noisy piecewise-smooth
dynamical system.
By applying the unified coloured noise approximation (UCNA), we obtained an analytical expression of the stationary probability density for the velocity. The white noise limit $\tau \rightarrow 0$ reproduces the exact results, e.g., 
see \cite{DEG05}. As the noise correlation time increases, 
the stationary density develops a delta peak, as particles become more 
and more stuck at $v = 0$. By varying $\tau$ a transition form sliding to sticking dynamics could be observed. By considering the equivalent two-dimensional system we were able to derive an asymptotic expression for the stationary density 
which is valid for large velocities and large noise amplitudes, far away from 
the stick-slip region. There was no obvious way to obtain a full analytic expression for the joint probability density $P(v,\eta)$ containing all the required matching conditions at $v = 0$ as detailed balance is violated. The latter 
has been clearly demonstrated by computing the non-vanishing stationary 
probability current.
 
Furthermore we studied the power spectral density numerically to obtain information about the velocity correlations, the corresponding correlation time, and the spectral gap of the underlying Fokker-Planck operator. Below a "critical value" one recovers the result for white noise limit. Increasing the noise correlation further, the full width at half maximum decreases, which is connected to a higher velocity correlation in the system. This decrease of the spectral width comes together with a change in shape of the power spectrum.
For low values of $\tau$ the power spectral density
is a Lorentzian while for values $\tau > 0.1$ the shape changes 
to a $\omega ^{-4}$ decay for intermediate frequencies. 

To complete our studies, we investigated the sliding and sticking time distribution as the time traces indicated a connection to intermittent dynamics. 
Results for the sticking time distribution were accessible via the exit time problem for an Ornstein-Uhlenbeck process with symmetric absorbing boundary conditions. For the sliding dynamics the related exit time problem with coloured noise and a constant drift is hard to tackle and we had to rely on 
simulations. For high noise correlation times a power law decay 
of the form $T^{-3/2}$ occurs, indicating a relation with
on-off intermittency.

The references \cite{GNO13} and \cite{GNO13a} provide probably the most
comprehensive experimental and theoretical
analysis of a device subjected to dry friction
and a nonequilibrium granular heat bath. The
corresponding theoretical considerations have
been based on a Boltzmann equation approach.
Results such as a ratchet effect induced
by geometric asymmetries and the localisation
of the velocity distribution are in accordance
with measurements. Given the sophisticated nature
of the underlying theoretical description, time
correlations and power spectra are not
accessible by analytic methods.

In our analysis we have addressed a simpler but related
theoretical model, using coloured noise instead
of a collision integral. There is no mathematical link
between both models, and the Boltzmann equation
and the dry friction model subjected to coloured
noise are fundamentally different. Nevertheless
we found various striking similarities. Time traces
of the coloured noise model are surprisingly similar to
those measured in experiments if the cases of
rare and frequent collision limits are compared
with large and small noise correlation time.
In addition, both models produce densities
with a singular component caused by the
discontinuous drift, a feature which is
common in a large class of piecewise-smooth
stochastic models, see e.g. \cite{BAU13}. Such a property can
be seen as an ubiquitous feature of stick slip
phenomena which is not restricted to a
particular theoretical or experimental
realisation. Within the analysis of
the dry friction model subjected to
coloured noise we were able to derive
an analytic expression for the weight
of the singular component, which is otherwise
hardly accessible (see e.g. as well \cite{BAU13}).
The analysis of the coloured noise model is
facilitated by a continuous control parameter, which has not
been available in the aforementioned more
realistic studies, where only the limiting cases
of frequent and rare collisions could be addressed.
We were able to identify a critical noise correlation time separating the white noise
regime from models where noise correlations
have a visible effect in the presence of
discontinuous drifts. Our model
allowed for a detailed analysis of nonequilibrium
currents and power spectra. In particular
the on-off intermittent characteristics is
a promising result which is tempting
to be checked experimentally.
In addition to the setup used
in \cite{GNO13a} a realisation along the lines of
\cite{GOO10} would allow to implement noise colour
quantitatively and thus would provide
a direct experimental comparison.

Apart from experimental confirmations the
coloured noise model is remarkable as well
from a plain theoretical perspective. In the
extended $(v,\eta)$ phase space the model is
described by a plain Fokker-Planck equation.
Because of the particular structure of diffusion and
discontinuous drift the two dimensional Gaussian
white noise model develops a singular stationary
density, proving that such a localisation phenomenon
is by no means a feature that requires more
complicated noise sources. Hence, features previously
found in Boltzmann equations can be certainly
captured by Fokker-Planck equations and simpler
stochastic models, which may be amenable for an
analytic treatment.



\end{document}